\documentclass[12pt]{article}

\newcommand{\hs}{\hspace{0.15mm}}

\usepackage{bm}
\usepackage{latexsym}
\usepackage[latin1]{inputenc}
\usepackage{amsfonts}
\usepackage{amssymb}
\usepackage{amsmath}

\numberwithin{equation}{section}

\begin{document}

\title{Can Hamiltonians be boundary observables in Parametrized Field Theories?}
\author{Tom\'{a}s Andrade, Donald Marolf   \\
{\it Department of Physics, UCSB, Santa Barbara, CA 93106, USA} \\ \\
C\'edric Deffayet  \\
{\it AstroParticule et Cosmologie, UMR7164-CNRS}, \\
{\it Universit\'e Denis Diderot-Paris 7, CEA, Observatoire de Paris,} \\
{\it 10 rue Alice Domon et L\'eonie Duquet,  F-75205 Paris Cedex 13, France} }

\maketitle

\begin{abstract}

It has been argued that holography in gravitational theories is related to the existence of a particularly useful Gauss Law that allows energy to be measured at the boundary.  The present work investigates the extent to which such Gauss Laws follow from diffeomorphism invariance. We study parametrized field theories, which are a class of diffeomorphism-invariant theories without gravity.  We find that the Hamiltonian for non-gravitational parametrized field theories vanishes on shell even in the presence of a boundary and under a variety of boundary conditions.  We conclude that such theories have no useful Gauss Law, consistent with the absence of holography.
\end{abstract}

\section{Introduction}

It is well known from the work or Arnowitt, Deser, and Misner (ADM) \cite{ADM} that the gravitational Hamiltonian reduces to a boundary term on-shell, and so vanishes identically for systems with no boundary in space. This non-linear gravitational Gauss law leads to a variety of phenomena, such as our ability to measure the total energy of a gravitating object from far away.  Of course, this includes the ability to measure the total energy in any black hole.

It has been recently argued \cite{DM1,DM2} that the gravitational Gauss law is also the key to understanding so-called ``holographic'' phenomena in gravitational physics, such as the AdS/CFT correspondence \cite{Juan} and unitarity in black hole evaporation.  In particular, the Gauss Law allows gravitational fields to store information in ways not possible in more familiar local field theories so that the algebra of boundary observables at infinity can be complete.  In some rough sense, this is because a large set of operators can be constructed through commutators of local fields at the boundary with the total energy. The critical properties for the arguments of \cite{DM1,DM2} are thus that the Hamiltonian i) becomes a pure boundary term on-shell, ii) can be considered a member of an appropriate (on-shell) ``algebra of boundary observables,'' and iii) generates non-trivial time translations along the boundary of this on-shell algebra.

Now, the existence of an on-shell algebra of boundary observables (property ii) is a rather mild assumption (see e.g. \cite{Rehren} for discussions of boundary observables in scalar field theories).  Furthermore, it is clear that property (i) is closely associated with diffeomorphism invariance.  In particular, diffeomorphism invariance means that the bulk part of the Hamiltonian generates a pure gauge transformation.  As a result, the bulk part of the Hamiltonian must commute with all gauge-invariant quantities.  In any on-shell (e.g., covariant phase space) formalism, this means that the bulk part of the Hamiltonian is constant over the space of solutions.  In some sense then the Hamiltonian will be a pure boundary term, which one would expect to generate time-translations.

On the other hand, diffeomorphism-invariance alone cannot be enough to yield features analogous to AdS/CFT.  The point here is that {\it any} local theory (e.g., a single free scalar field) can be written in diffeomorphism-invariant form through a process known as parametrization\footnote{See  \cite{ADM,ADM2,ADM3,ADM4,Dirac,Kuchar 1-4,Isham:1984sb} for earlier discussions.}.  But it is clear that free (unparametrized) scalar fields are not in themselves holographic since time evolution mixes boundary observables at any one time $t$ with independent bulk observables (say, those space-like separated from the cut of the boundary defined by the time $t$).   As a result, boundary observables at one time cannot generally be written in terms of boundary observables at any other time.  In the language of \cite{DM1}, we may say that the algebra of boundary observables does not evolve unitarily in time and thus that the theory does not exhibit `boundary unitarity.'  In contrast, this form of unitarity of the boundary theory is a key property of AdS/CFT.

It therefore appears that at least one of the properties (i,ii,iii) must fail in general for non-gravitational parametrized theories\footnote{\label{foot} The astute reader may note that one may rewrite the above scalar field theory, and indeed any local field theory, in a form that does satisfy (i), (ii), and (iii) by introducing a new field $\varphi$ that satisfies the constraint $\nabla^2 \varphi = 4 \pi \rho$ at each time $t$, where $\nabla^2$ is the Laplacian on a constant $t$ surface and $\rho$ is the energy density.  It is then clear that the Hamiltonian can be written in terms of the normal derivatives of $\varphi$ at the boundary.  But such theories are not holographic in an interesting way, and are distinguished by the fact that the algebra of boundary observables is non-local; i.e., that equal time commutators at separated points on the boundary do not vanish.  In this work, we implicitly confine discussion to theories for which the algebra of boundary observables is {\it local}, by which we mean that the boundary can be foliated by a set of cuts ${\cal C}_t$ and that, given any open cover $U_i$ of any cut, this algebra a) is generated by the union of sub-algebras associated with each $U_i$ and b) observables commute if they are associated with disjoint sets $U_i, U_j$.}.  We see two ways in which this can happen.  First, although the bulk part of the Hamiltonian vanishes on shell in the sense noted above, it might be that for technical reasons the Hamiltonian does not reduce to a well-defined boundary observable, so that property (ii) fails.  The other option is that the Hamiltonian is actually trivial, or at least acts trivially on any boundary observables.

That one of these properties fails for non-gravitational parametrized theories should be no surprise, as the particular form of the gravitational constraints are known to play a key role in constructing the ADM energy as a boundary term \cite{ADM} in general relativity and its analogues in other theories of gravity \cite{DT}.  In addition, as we will review in section \ref{Action principle and boundary conditions}, it is clear from the analyses of \cite{ADM2,Kuchar 1-4} that the Hamiltonian vanishes for parametrized scalar field theories satisfying the simplest boundary conditions.
What remains is to investigate more complicated boundary conditions in detail.

One step in this direction comes from \cite{peierls argument}, which studied systems with time-translation-invariant boundary conditions specified by scalars and a vielbein and which defined the algebra of boundary observables using the Peierls bracket \cite{Peierls}.  In such settings, it was shown that the associated symmetry generator of any local diffeomorphism-invariant theory can be constructed from a so-called ``boundary stress tensor'' $T_{ab}$ (analogous to that defined in \cite{HS,BK}) constructed by varying the action with respect to the boundary condition placed on the metric (or vielbein) field.  Suppose then that we have a system which contains only scalars (and no metric) but which otherwise satisfies the conditions of \cite{peierls argument}.  Then it may be extended to a system with metric, but for which the action $S$ remains independent of the metric.  This means that the boundary stress tensor vanishes identically ($T_{ab} =0$), and so does the Hamiltonian.

As we review below, the so-called differential map of a parametrized field theory plays the role of a vielbein field. As a result, any parametrized field theory may be regarded as a theory of scalars.  It then follows from \cite{peierls argument} as above that the Hamiltonian, at least as defined by the Peierls bracket, must vanish for any time-translation-invariant boundary conditions. Thus property (iii) fails in such settings.

Our purpose here is to explore Hamiltonians for non-gravitational parametrized systems\footnote{Below, we refer to such systems as simply ``parametrized theories,'' with the term "non-gravitational" being left implicit.  We have in mind some non-gravitational local field theory that has been parametrized by hand, in analogy to the process described for scalar field theories below.   The reader should note that our use of this term differs from that of e.g. ADM \cite{ADM}, who describe general relativity as an ``already parametrized theory."} with boundaries in more detail.  In particular, it is natural to ask two questions:  First, since the Peierls bracket is less familiar to most readers than either canonical methods or covariant methods based on the symplectic structure, one may ask if the above result can be simply an artifact of the Peierls formalism.  Certainly, it is of interest to understand the result using either canonical or covariant phase space methods as well.  Second, one may ask whether more interesting results might be obtained in some context where the boundary conditions do not satisfy time-translation invariance.

We investigate such issues below. We begin with a brief discussion of parametrized scalar fields and possible boundary conditions in section \ref{PFT}.  This discussion is covariant, and addresses the relevant boundary terms in the action. The boundary conditions we consider for the dynamical scalar $(\phi)$ are Dirichlet, Neumann, or Robin.  The case of Robin boundary conditions will be of particular interest since then the natural form of the Hamiltonian for the unparametrized theory already contains an explicit boundary term.  The boundary conditions we consider for the diffeomorphism-scalars ($X^\alpha$, which are introduced to parametrize the system) are termed minimal, velocity-fixed, and fully-fixed and are explained below.

We then review the canonical phase space in section \ref{can}, following \cite{ADM2} and \cite{Isham:1984sb}.  This analysis clearly shows that the Hamiltonian vanishes on shell when the dynamical scalar $\phi$ satisfies Dirichlet or Neumann boundary conditions, for either minimal or velocity-fixed boundary conditions on the $X^\alpha$.  However, for fully fixed boundary conditions on the $X^\alpha$ or for Robin boundary conditions $\phi$ it is less clear whether a good canonical formalism exists.

We therefore turn to covariant phase space methods in section \ref{cov}.  Such methods are well-defined and lead to useful insights even for Robin boundary conditions on $\phi$ and for fully-fixed boundary conditions on $X^\alpha$.  For minimal and velocity-fixed boundary conditions on the $X^\alpha$ we find that time-translations are degenerate directions of the symplectic structure, so any generator is a constant on-shell and may be taken to vanish identically. The fully-fixed boundary conditions are more subtle as strict time-translations no longer preserve the boundary conditions.  However, we explore a large family of possible notions of modified time translations which preserve the boundary conditions.  In each case we find that either a) the modified transformation is again a degenerate direction and the Hamiltonian vanishes, b) for technical reasons we are unable to show that the associated Hamiltonian is a member of the algebra of boundary observables, or c) the modified transformation has no well-defined generator as it defines a non-Hamiltonian vector field on the covariant phase space; i.e., the equations that would define the desired generator are not integrable.

\section{Parametrized scalar with boundary}
\label{PFT}

We now briefly review the parametrized scalar field. Our discussion generally follows that of \cite{Torre}, but adds the elements required for a full treatment of boundaries. See also \cite{ADM2} and \cite{Kuchar 1-4} for earlier discussions.  For simplicity, we consider the boundary to be at a finite location.  However,  for the usual reasons the treatment of asymptotically-AdS boundaries will be similar.

The action
\begin{equation}\label{unparam action}
    S_0 = -\frac{1}{2} \int_M d^4 X \sqrt{g} g^{\alpha \beta} \partial_\alpha \phi \partial_\beta \phi,
\end{equation}
for an unparametrized scalar field $\phi$ on a spacetime with metric $g_{\alpha \beta}$ and coordinates $X^\alpha$ is suitable for Dirichlet boundary conditions ($\phi$ fixed on the boundary) or what one might call ``pure Neumann'' boundary conditions ($\partial_\rho \phi =0$, where $\partial_\rho$ denotes the derivative along the unit normal $\rho^\alpha$) at any finite boundary of the domain of the $X^\alpha$.  By saying that the action is ``suitable,'' we mean that the variation of the action vanishes on-shell precisely when both the boundary conditions and the Euler-Lagrange equations of motion hold, and in particular that all boundary terms in the variation vanish.  For more general Robin boundary conditions of the form
\begin{equation}\label{robin bcs}
    \partial_\rho \phi = \alpha(X) \phi + \beta(X),
\end{equation}
where $\alpha$ and $\beta$ are arbitrary functions of $X$, a suitable action is
\begin{equation}\label{S robin}
	S_{Robin} = S_0 + S_{Robin,bndy} = S_0 + \int_{\partial M} \sqrt{h} [  \phi \partial_\rho \phi   - \frac{\alpha}{2}  \phi^2 ],
\end{equation}
where $h$ is the determinant of the induced metric on the boundary. Below, we will assume that any boundary can be described as the solution of some equation $F(X)=0$.

We may parametrize the actions (\ref{unparam action}) and (\ref{S robin}) by enlarging the configuration space to include diffeomorphisms of $M$ to itself, in addition to the degree of freedom described by $\phi$. It is convenient to work with two copies of $M$, denoted by $M^\alpha$ (target space) and $M^\mu$ (the coordinate space). Then, we can think of $X$ as a (smooth) map
\begin{equation*}
    X: M^\mu \rightarrow M^\alpha \,\,\,\,\,\,\,\,\,\, X^\alpha = X^\alpha(y^\mu).
\end{equation*}
Below, we require this map to be bijective, and in particular to map $\partial M^\mu$ to $\partial M^\alpha$.  However, a generalization along the lines of \cite{partial systems} should also be possible without this restriction.

The parametrized action is given by pulling back the fields to $M^\mu$, $\phi(y) = \phi(X(y))$, $G_{\mu \nu} = X^\alpha_\mu X^\beta_\nu g_{\alpha \beta} $, where we have introduced the differential map $X^\alpha _\mu \equiv \frac{\partial X^\alpha}{\partial y^\mu}$. For Dirichlet or Neumann boundary conditions on $\phi$, the action thus reads
\begin{equation}\label{param action}
    S_0^{P}[\phi, X] = -\frac{1}{2} \int_{M^\mu} d^4 y \sqrt{G} G^{\mu \nu} \partial_\mu \phi \partial_\nu \phi   = -\frac{1}{2} \int_{M^\mu} d^4 y  \bigg| \frac{\partial X}{\partial y} \bigg| \sqrt{g} g^{\alpha \beta} X^\mu _\alpha X^\nu _\beta \partial_\mu \phi \partial_\nu \phi.
\end{equation}
In the Robin case, this is supplemented by a similar pull-back $S_{Robin,bndy}^P$ of the boundary term $S_{Robin,bndy}$.  We define $S_{Robin}^{P} = S_0^{P} + S_{Robin,bndy}^{P}$.  Note that the differential map $X_\mu^\alpha$ plays the role of a vielbein, and can thus be used to formulate any parametrized field theory (e.g., even one a priori involving vector fields) in terms of spacetime scalars.

Since our parametrized actions are simply the pull backs of good variational principles defined by $S_0, S_{Robin}$, they must also yield good variational principles under the pull-back of the appropriate boundary conditions; i.e., when $F(X) =0$ on the boundary of $M^\mu$ and when $\phi$ satisfies the appropriate Dirichlet, Neumann, or Robin boundary condition (\ref{robin bcs}). However, another way to arrive at this key result is to note that, since a variation of the $X^\alpha$ fields acts as a diffeomorphism in the unparametrized theory, the on-shell variation of $S_0^P$ with respect to $X^\delta$ must be $\int_{\partial M} \sqrt{h}  T_{\alpha \beta} \rho^\beta \delta X^\alpha$ where $T_{\alpha \beta}$ is the stress tensor
\begin{equation}\label{Tab}
    T_{\alpha \beta} = \partial_\alpha \phi \partial_\beta \phi + g_{\alpha \beta} L,
\end{equation}
with $L$ defined by (\ref{unparam action}).   Indeed, an explicit calculation (see appendix \ref{detailed variation DN}) yields
\begin{equation}\label{delta S DN}
    \delta S_0^P = \int_{\partial M^\mu} \sqrt{h} \left \{
   T_{\alpha \beta} \rho^\beta \delta X^\alpha - \partial_\rho \phi \delta \phi    \right\}  ,
\end{equation}
where now $\partial_\rho \phi \equiv \rho^\alpha \left( \frac{\partial X^\alpha}{\partial y^\mu}\right)^{-1} \partial_\mu \phi$.  Using
$\partial_\alpha F \propto \rho_\alpha$ and $\rho_\alpha \delta X^\alpha =0$ on $\partial M^\mu$, it is clear that (\ref{delta S DN}) vanishes when $\phi$ satisfies the pure Neumann boundary condition ($\partial_\rho \phi =0$).   The result is also clear for Dirichlet boundary conditions ($\phi = \phi_0(X)$) using the fact that $\delta X^\alpha$ is tangent to the boundary and thus that $\delta X^\alpha \partial_\alpha \phi =0$ on $\partial M^\mu$.   In the Robin case, one may explicitly check that (\ref{delta S DN}) is canceled by the variation of $S_{Robin,bndy}^P$, see appendix \ref{detailed variation DN} for details.
 For reasons discussed below, we call $F(X)=0$ the minimal boundary condition on the $X^\alpha$.

By construction, the actions $S_0^{P}, S_{Robin}^{P}$  are invariant under diffeomorphisms $y'^\mu = y^\mu - \epsilon^\mu(y)$ of our manifold with boundary, provided that all the fields transform as scalars:
\begin{equation}\label{gauge syms}
    \delta \phi = \epsilon^\mu \partial_\mu \phi \,\,\,\,\,\,\,\,\,\,  \delta X^\alpha = \epsilon^\mu \partial_\mu X^\alpha .
\end{equation}
Note that the metric $g_{\alpha \beta}$ is also a scalar, so $\delta g_{\alpha \beta} = \partial_\gamma g_{\alpha \beta} \delta X^\gamma$. Indeed, it can be readily verified that the variation of (\ref{param action}) under (\ref{gauge syms}) is
\begin{equation*}
    \delta S = \int_{M^\mu} \partial_\mu ( {\cal L} \epsilon^\mu ),
\end{equation*}
which vanishes so long as $\epsilon^\mu \rho_\mu = 0$, where $\rho_\mu = \partial_\mu X^\alpha \rho_\alpha$ is the pull back to $M^\mu$ of the unit normal to the boundary.

Because diffeomorphisms that act non-trivially on the boundary define {\it local} symmetries of our actions,  we expect that all such diffeomorphisms are pure gauge (so long as they map $\partial M^\mu$ to itself).  This differs from familiar cases involving gravity, where diffeomorphisms that act non-trivially on the boundary define useful asymptotic symmetries and only diffeomorphisms that act trivially are pure gauge.  In particular, if boundary diffeomorphisms are pure gauge one expects the resulting algebra of boundary observables not to be local in the sense of footnote \ref{foot}.  It is thus natural to ask if other boundary conditions would give better analogues of the gravitational setting.  To investigate this question, we will study two additional classes of boundary conditions below:
\begin{itemize}
\item{\bf Velocity-fixed boundary conditions:} Choose some scalar field $t$ on the boundary for which $dt$ does not vanish.  Also choose some non-vanishing vector field $t^\mu$ on the boundary for which $t^\mu \partial_\mu t =1$.  Then, in addition to setting $F(X) =0$, we may fix $\dot{X}^\alpha :=  t^\mu \partial_\mu X^\alpha$ to be constants on $\partial M^\mu$ .  In this case, time translations define global, but not local symmetries of the system.  We may ask if the symmetry generator is non-trivial.

\item{\bf Fully-fixed boundary conditions:} Choose some diffeomorphism $f$ from $\partial M^\mu$ to ${\partial M}^\alpha$.  Then we may fix $X^\alpha = f^\alpha$ on $\partial M^\mu$ (in addition, of course, to $F(X)=0$).   Note that this boundary condition imposes a well-defined causal structure on the boundary such that two boundary points are causally related for any solution if and only if they are causally related in the corresponding unparametrized theory.  As a result, the algebra of boundary observables will necessarily be local in the sense of footnote \ref{foot}.  While time translations are no longer symmetries of the system, we may still ask if there is a (necessarily time-dependent) Hamiltonian which generates some modified notion of time-translation.
\end{itemize}
Since both of these boundary conditions require $F(X)=0$ (in addition to other conditions) on $\partial M^\mu$, it is clear that our actions $S_0^P, S_{Robin}^P$ again give good variational principles for these new boundary conditions for $X^\alpha$, so long as the appropriate Dirichlet, Neumann, or Robin boundary condition is also imposed on $\phi$.

\section{Canonical Formalism}
\label{can}

We begin our investigation of the Hamiltonian with the canonical formalism, as this is the most familiar framework for studying such issues.  The canonical formalism for parametrized field theories on manifolds without boundary was studied in detail in \cite{ADM2}, \cite{Kuchar 1-4}, and \cite{Isham:1984sb} building on \cite{ADM,Dirac}. We review this construction below, making explicit the issues associated with manifolds with boundaries.  Our discussion generally follows that of \cite{Kuchar 1-4}.

\subsection{Geometry review}
\label{georev}

Consider the manifold $M$ associated with the unparametrized theory (\ref{unparam action}) and the associated metric $g$. Let us assume that $M$ has the topology of a cylinder $M = \Sigma \times {\mathbb R}$, where $\Sigma$ is the spatial manifold. We are interested in the case where $M$ has a time-like boundary $\partial M$. It is convenient for the canonical formalism to describe the geometry as a foliation given by a smooth family of embeddings $\Sigma_t \rightarrow M$ where $t$ is a smooth parameter that we will call time.  As in section (\ref{PFT}), the coordinates on $M$ are denoted $X^\alpha$.  Coordinates on $\Sigma$ will be $x^a = (r,x^m)$ so that the foliation is $X^\alpha = X^\alpha(t,x^a)$ and we may take the $y^\mu$ of section (\ref{PFT}) to be $y^\mu=(t,x^a)$. It is convenient to choose coordinates such that $r=r_0$ is a constant on $\partial M^\mu$.   As a result, diffeomorphisms generated by the vector field $\partial_t$ map $\partial M$ to itself. Coordinates on $\partial M^\mu$ will be denoted by $y^A = (\hat t,\hat x^m)$.  Here $(\hat t, \hat x^m)$ are simply the restrictions of $(t,x^m)$ to the boundary, but the hats $(\hat{})$ will help to avoid certain ambiguities below.\\

Note that, since it maps vectors in $M$ to vectors in $\Sigma$,  the differential map $\partial_a X^\alpha \equiv  X^\alpha_a$ can be thought of as a projector into each slice $\Sigma$. Furthermore, there exists a one form $n = n_\alpha dX^\alpha$ that is orthogonal to $\Sigma$,
\begin{equation}\label{normal}
    n_\alpha X^\alpha_a = 0  ,
\end{equation}
which can be taken to be normalized everywhere
\begin{equation}\label{n normalization}
    n_\alpha n^\alpha = -1  .
\end{equation}
We can therefore decompose every vector in $M$ into its tangential and normal components with respect to $\Sigma$,
\begin{equation}\label{decomposition}
    V^\alpha = V n^\alpha + X^\alpha_a V^a \,\,\,\,\,\,\,\,\,\,\,\,\, V = -n_\alpha V^\alpha \,\,\,\,\,\,\,\, V^a = X_\alpha^a V^\alpha   .
\end{equation}
In particular, it is useful to define the deformation vector $\dot{X}^\alpha =\partial_t X^\alpha \Big|_{x^a}$ which relates two infinitesimally close slices. It can  be decomposed as
\begin{equation}\label{decomposition of N}
    \dot{X}^\alpha = N n^\alpha + X^\alpha_a N^a \,\,\,\,\,\,\,\,\,\,\,\,\, N = -n_\alpha N^\alpha \,\,\,\,\,\,\,\, N^a = X_\alpha^a N^\alpha ,
\end{equation}
where $N$ and $N^a$ are called lapse and shift. The induced metric on $\Sigma$ is
\begin{equation}\label{induced metric}
    \gamma_{a b} = X^\alpha_a g_{\alpha \beta} X^\beta_b  .
\end{equation}

Similarly, we also define the unit normal $\rho_\alpha$ to the time-like boundary by $\rho_\alpha X^\alpha_A = 0$ and $\rho^\alpha \rho_\alpha = +1$. It is useful to express it in terms of the basis $n_\alpha$, $X^\alpha_a$,
\begin{equation}\label{radial normal}
    \rho_\alpha = \rho n_\alpha + X^a_\alpha \rho_a .
\end{equation}
Note that since $\partial_t$ preserves $\partial M$, we have $0 = \dot{F} \propto \rho_\alpha \dot{X}^\alpha$ and similarly $X^\alpha_m \rho_\alpha = 0$.  Together with the normalization condition $\rho^\alpha \rho_\alpha = +1$, these facts imply \begin{equation*}
    \rho_\alpha = \frac{\rho_r}{N}(N^r n_\alpha + N X^r_\alpha ).
\end{equation*}
In terms of the induced metric $h_{AB} = X^\alpha_A g_{\alpha \beta} X^\beta_B$
 on the boundary, for which $\sqrt{g} = \rho_\alpha X^\alpha _r \sqrt{h} = \rho_r \sqrt{h}$, we have
\begin{equation}\label{radial normal explicit}
    \rho_\alpha = \frac{\sqrt{\gamma}}{\sqrt{h}}(N^r n_\alpha + N X^r_\alpha ).
\end{equation}

\subsection{Dirichlet and Pure Neumann Boundary Conditions}
\label{Action principle and boundary conditions}

We now wish to use the technology introduced in section (\ref{georev}) to compute the canonical Hamiltonian, thus writing the action  in canonical form.
We begin with the action (\ref{param action}), appropriate for Dirichlet or pure Neumann boundary conditions on $\phi$ and save the Robin case for section \ref{Robin boundary conditions}.

Pulling back the field $\phi$ to $\Sigma$ as $\phi(X) = \phi(X(t,x^a))$ yields
\begin{equation}\label{gradient decomp}
    \partial_\alpha \phi = \phi_{, \bot} n_\alpha + X^a_\alpha \phi_{,a}  \,\,\,\,\,\,\,\,\,\,\,\,\, \phi_{, \bot} = -n^\alpha \phi_{,\alpha} \,\,\,\,\,\,\,\, \phi_{,a} = X^\alpha_a \phi_{,\alpha}  ,
\end{equation}
and thus
\begin{equation}\label{param lagrangian}
    S^P_0 = \int d^4 x N \sqrt{\gamma} \left \{ \frac{1}{2} [ (\phi_\bot)^2 - \gamma^{ab} \phi_{,a} \phi_{,b}  ]  - V(\phi)   \right \} .
\end{equation}
We now compute the conjugate momenta
\begin{equation}\label{pi}
    \pi = \frac{\delta {\cal L}}{\delta \dot{\phi}} = - \gamma^{1/2} \phi_{,\bot}, \ \ \
    P_\alpha = \frac{\delta {\cal L}}{\delta \dot{X}^\alpha} = - H^\phi_{\alpha} ,
\end{equation}
\noindent where ${\cal L}$ is the Lagrangian density ${\cal L} = \sqrt{g} L$ and
\begin{equation}\label{H}
    H^\phi_\alpha = - n_\alpha H^{\phi} + X_\alpha^a H^\phi_a
\end{equation}
\begin{equation}\label{H alpha}
    H^\phi = \frac{1}{2}\left [ \frac{\pi^2}{\gamma^{1/2}} + \gamma^{1/2}\gamma^{ab}\phi_{,a} \phi_{,b} \right ]  + \gamma^{1/2} V(\phi) \,\,\,\,\,\,\,\,\,\,\,\,\,\,\, H^\phi_a =\pi \phi_{,a} .
\end{equation}

Let us now calculate the canonical Hamiltonian density in the parametrized theory. First, we note that the canonical Hamiltonian density $H_{can}^{un}$ in the {\it unparametrized} theory may be written in the familiar form
\begin{equation}\label{H can unparam}
    H^{un}_{can}  = N H^\phi + N^a H_a^\phi = \dot{X}^\alpha H^\phi_\alpha \,\,\,\,\,\,\, ,
\end{equation}
but where the lapse and the shift are not dynamical variables. Instead, they are fixed in terms of the background metric. When we parametrize the theory, the embedding variables become canonical variables.  Thus
\begin{equation}\label{H can param}
    H^P_{can} = P_\alpha \dot{X}^\alpha + \pi \dot{\phi} - {\cal L} = P_\alpha \dot{X}^\alpha + H_{can} = 0,
\end{equation}
where we have used (\ref{pi}) to show that the final result vanishes.

Equations (\ref{H}) and (\ref{H alpha}) imply that (\ref{pi}) cannot be used to express $\dot{X}^\alpha$ in terms of $P_\alpha$.  We must therefore add appropriate constraints and Lagrange multipliers $\lambda^\alpha$ to the canonical action.  As in \cite{ADM2,Kuchar 1-4}, the result is
\begin{equation}\label{S scalar parametrized}
    S_{0,can}^P[X^\alpha, P_\alpha, \phi, \pi; \lambda^\alpha] = \int dt d^d x \bigg [ \pi \dot{\phi} + P_\alpha \dot{X}^\alpha  - \lambda^\alpha(P_\alpha + H^\phi_\alpha ) \bigg ]  .
\end{equation}
Since it is equal on-shell to $S_0^P$, this action is suitable for Dirichlet or pure Neumann boundary conditions on $\phi$ and for minimal, velocity-fixed, or fully-fixed boundary conditions on the $X^\alpha$.

At the formal level, it would appear that each of these cases leads to a good canonical formalism and that the canonical Hamiltonian vanishes on-shell.  In familiar settings, the fact that the canonical formalism quantities requires observables to commute with all constraints means that the observable algebra can be taken to be an on-shell algebra.  In the present case, this would mean that all observables must commute with the Hamiltonian.  Thus for velocity-fixed boundary conditions, where a priori one might have expected time translations to define asymptotic symmetries, and even for fully-fixed boundary conditions, where one might have expected that time-translations are not symmetries at all, all observables would be invariant under time-translations.  In this sense, time-translations would remain pure gauge even for these more restrictive boundary conditions.   However, there are clearly subtleties to consider since the Hamiltonian
(\ref{H can param}) appears to act non-trivially on the $X^\alpha$ at the boundary and thus fails to preserve the fully-fixed boundary conditions.  Rather than address the associated technical details here, for the moment we merely take the above analysis as suggestive.  It will be more convenient to analyze the details using the covariant phase space approach of section \ref{cov}.  In particular, the fully-fixed boundary conditions will be treated in section \ref{FF}.

\subsection{Robin boundary conditions}
\label{Robin boundary conditions}

As noted earlier, the case of Robin boundary conditions may be expected to be the most interesting due to the presence of the explicit boundary term $S_{Robin,bndy}$ in the unparametrized action.  Here we briefly investigate the consequences for the canonical formalism.  The story is in direct parallel with the Dirichlet/pure Neumann case, except that the boundary metric $h$ depends on $\dot{X}^\alpha$.  As a result, at least for minimal boundary conditions on the $X^\alpha$, the expression for $P_\alpha$ is modified\footnote{One may use the boundary condition (\ref{robin bcs}) to eliminate the other velocities that appear in the boundary term so that $\pi$ remains unmodified}. Thus, the canonical momenta read
\begin{equation}\label{P robin}
    \pi = - \gamma^{1/2}\phi_{,\bot}  , \ \ \
    P_\alpha = - H^\phi_{\alpha} +   \sqrt{h} X_\alpha^{\hat t} [  \phi \partial_\rho \phi   - \frac{\alpha}{2}  \phi^2 ] \delta(r,r_0)  ,
\end{equation}
where the delta function takes into account the fact that the second term in (\ref{P robin}) only contributes at the boundary. Again, an explicit calculation shows that the Hamiltonian density $H_{can} = p_i \dot{q}^i - {\cal L}$ vanishes, and even the boundary term in the action (\ref{S robin}) is exactly canceled by the boundary term in the momenta (\ref{P robin}).  This last point can be seen by noting that $\dot{X}^\alpha X_\alpha^{\hat t} = 1$.  Once more, we find that the canonical Hamiltonian vanishes.

However, it is difficult to find a good canonical action for this system.  To illustrate the point, suppose that we simply add  the definition of the $P_\alpha$ to the action as constraints.  This yields
\begin{equation}\label{S param robin}
    S = \int dt \int_{\Sigma} d^3 x  \{ \pi \dot{\phi} + P_\alpha \dot{X}^\alpha - \lambda^\alpha [P_\alpha + H^\phi_{\alpha}] \}
    -  \int dt \int_{\partial \Sigma} d^3x \lambda ^\alpha \sqrt{h} X^{\hat t}_\alpha  [  \phi \partial_\rho \phi   - \frac{\alpha}{2}  \phi^2 ] .
\end{equation}
As the reader will note, the boundary term in (\ref{S param robin}) still depends on velocities which have not been written in terms of momenta.  One might like to use (\ref{P robin}) to replace velocities on the boundary by momenta, but it is not clear to us how this might be usefully achieved.  The greatest problem appears to be the presence of the delta-function on the right-hand side of (\ref{P robin}).  If one imagined a discretized version of this theory in which the delta-function were replaced by a Kronecker delta, there would be no obstacle.  This observation raises concerns about the extent to which the canonical formalism is well-defined for the parametrized scalar with Robin boundary conditions.  We therefore turn to a covariant phase space analysis, in which it will be possible to check directly whether the resulting phase space is well-defined.  For related reasons, we postpone discussion of velocity-fixed and fully-fixed boundary conditions to section \ref{cov} as well.

\section{Covariant Phase Spaces}
\label{cov}

We now clarify various issues raised in the canonical approach by analyzing our parametrized field theory with boundary using covariant phase space technology.  The key element in such constructions is the symplectic structure, which we will need to define so that it is conserved under the appropriate boundary conditions.  In the formalism of e.g. \cite{LW}, this property is not guaranteed a priori and one must attempt to correct any non-conservation by adding an appropriate boundary term to the symplectic structure.
In particular, changes of boundary conditions must generically be accompanied by the addition of further boundary terms to the symplectic structure.  A prescription for linking the symplectic structure to boundary terms in a variational principle was introduced in \cite{CM,AC}.  We therefore briefly review both the covariant phase space formalism and the prescription of \cite{CM,AC} before beginning specific calculations below.

Suppose that our variational principle is given, up to possible boundary terms, by the integral of  Lagrange density ${\bf L}$.  Here we take ${\bf L}$ to be a $d$-form in $d$-dimensional spacetime.  Then the construction of the covariant phase space begins with the computation of the Euler-Lagrange equations.  Specifically, one defines the symplectic potential $d-1$ form $\boldsymbol \theta[\delta]$ by writing
\begin{equation}
\label{dL}
\delta {\bf L} = \delta_{EL} {\bf L}  + {\bf d} \boldsymbol{\theta}[\delta],
\end{equation}
where $\delta_{EL}$ is the so-called ``Euler-Lagrange variation," meaning that it is precisely the combination of derivatives with respect to coordinates and velocities (and, more generally, higher time-derivatives) that yields the Euler-Lagrange equations for the action (or their higher-derivative equivalents).  The fact that $\delta_{EL} {\bf L}$ differs from $\delta {\bf L}$ by a total derivative is the statement that the variation of the action yields the Euler-Lagrange equations after integrations by parts and up to boundary terms.  Of course, (\ref{dL}) defines ${\boldsymbol \theta}[\delta]$ only up to the addition of an exact form (a ``boundary term'').  The notation $\boldsymbol{\theta}[\delta]$ indicates that $\boldsymbol \theta$ takes as an argument a variation $\delta$ of the full set of fields.

The next step is to compute the symplectic $d-1$ form $\boldsymbol \omega = \delta_{[12]} \boldsymbol \theta := \delta_1  \boldsymbol \theta[\delta_2] - \delta_2 \boldsymbol \theta[\delta_1]$, where we have introduced the symbol $\delta_{[12]}$ to denote this anti-symmetric variation.  The symplectic structure is then the integral
\begin{equation}
\Omega = \int_\Sigma \boldsymbol {\omega}
\end{equation}
over any complete hypersurface $\Sigma$ (i.e., for which $\partial \Sigma \subset \partial M$). Shifting $\boldsymbol \theta$ by an exact form ${\bf dB}$ also shifts $\boldsymbol \omega$ by the exact form ${\bf d \delta_{[12]} B}$,  and so shifts $\Omega$ by a boundary term $\int_{\partial \Sigma} \delta_{[12]} {\bf B}$.

Suppose that one may choose the exact form ${\bf B}$ so that $\boldsymbol \theta$ vanishes when pulled-back to $\partial M$.  Then $\boldsymbol \omega$ also vanishes on $\partial M$, and $\Omega$ is independent of the choice of surface $\Sigma$; i.e., $\Omega$ is conserved.  Now consider a region $R \subset M$ bounded by two complete hypersurfaces $\Sigma_1, \Sigma_2$.  Then we have
\begin{equation}
\delta \int_R {\bf L} \approx \int_{\partial R} \boldsymbol \theta,
\end{equation}
where $\approx$ denotes equality on-shell.  Of course, the part of the left-hand-side associated with $\partial M$ vanishes.  In this sense, the boundary term in $\delta S$ vanishes locally on $\partial M$. So finding a choice of $\boldsymbol \theta$ that vanishes on $\partial M$ is equivalent to finding a variational principle in which the boundary terms of $\delta S$ vanish locally on $\partial M$, and such a choice guarantees conservation of $\Omega$.

The observation of \cite{CM} is that finding such a variational principle often involves adding explicit boundary terms to the action.  This can be viewed as shifting ${\bf L}$ by an exact form, with an associated change of $\boldsymbol \theta$. It is clear from the above argument that the {\it shifted} symplectic potential should vanish on $\partial M$.  Suppose in particular that $S = \int_M {\bf L}_{bulk} + \int_{\partial M} {\bf L}_{bndy}$ for some $d-1$ form ${\bf L}_{bndy}$.  Let us define $\boldsymbol{\theta}_{bulk}$, $\boldsymbol{\theta}_{bndy}$, and $\boldsymbol \theta$ by
\begin{eqnarray}
\label{CMdefs}
\delta {\bf L}_{bulk} &=& \delta_{EL} {\bf L}_{bulk}  + {\bf d} \boldsymbol{\theta}_{bulk}[\delta], \cr
\delta {\bf L}_{bndy} &=& \delta_{EL} {\bf L}_{bndy}  + {\bf d} \boldsymbol{\theta}_{bndy}[\delta], \cr
{\rm and} \ \boldsymbol \theta &=& \boldsymbol \theta_{bulk} - d \boldsymbol \theta_{bndy}.
\end{eqnarray}
Then the boundary terms in $\delta S$ will again vanish locally on $\partial M$ if
\begin{equation}
\boldsymbol \theta_{bulk} + \delta_{EL} L_{bndy} =0 \ {\rm on} \ \partial M.
\end{equation}
While this need not imply that $\boldsymbol \theta =0$ on $\partial M$, it does imply that $\boldsymbol \omega$ (the anti-symmetric variation of $\boldsymbol \theta$) vanishes when pulled back to $\partial M$.  The key point here is that since $ \delta_{EL} {\bf L}_{bndy}  + {\bf d} \boldsymbol{\theta}_{bndy}[\delta]$ is a total variation, its anti-symmetric variation must vanish.  Thus
\begin{equation}
\boldsymbol \omega_{bulk} - d \boldsymbol \omega_{bndy} =0 \ {\rm on} \ \partial M.
\end{equation}

In the context of parametrized field theories, these observations give a prescription for defining conserved symplectic structures from the actions $S_0^P, S_{Robin}^P$.  Although we have already studied the Dirichlet/pure Neumann case using canonical methods, it is useful to address such boundary conditions again in the covariant setting before moving on to the more complicated Robin case.  This will also allow us to understand the subtleties of fully-fixed boundary conditions on the $X^\alpha$ and to analyze more fully which diffeomorphisms are pure gauge.  We do so in section \ref{DNcov} below.  We then add the appropriate boundary terms to the symplectic structure for Robin boundary conditions in section \ref{Rcov}.

\subsection{The Dirichlet/pure Neumann case}
\label{DNcov}

In order to compute the symplectic structure for the action (\ref{param action}), we find it convenient to write the symplectic potential $d-1$ form ${\boldsymbol \theta}$ as the anti-symmetric symbol contracted with a current density $\tilde \theta^\mu$, and to similarly replace the symplectic $d$-form ${\boldsymbol \omega}$ by $\tilde \omega^\mu$.  As a result, this current density is defined by  $\delta {\cal L} = (eoms)\delta \Phi + \partial_\mu \tilde {\theta}^\mu$, where ${\cal L}$ is the Lagrangian viewed as a scalar density. For the action (\ref{param action}), we obtain
\begin{equation}\label{potential}
    \tilde{\theta}^\mu(\delta_1 X , \delta_1 \phi ) = \sqrt{G} \theta^\mu , \,\,\,\,\,\,\,\,\,\,\, \theta^\mu =  T_\alpha^\mu \delta_1 X^\alpha - \pi^\mu \delta_1 \phi,
\end{equation}
where $T^\alpha _\mu = X^\alpha _\nu G^{\nu \lambda} T_{\lambda \mu}$, $\pi^\mu = G^{\mu \nu} \partial_\nu \phi$ and the stress tensor $T_{\mu \nu}$ is the pull-back of (\ref{Tab}).
The densitized symplectic current is thus
\begin{equation}\label{omega}
    \tilde{\omega}^\mu = \delta_2 \sqrt{G} \theta^\mu[\delta_1] + \sqrt{G}[ \delta_2 T_\alpha^\mu \delta_1 X^\alpha - \delta_2 \pi^\mu \delta_1 \phi   ] - (1 \leftrightarrow 2)
\end{equation}

By the usual arguments, one can show that $\partial_\mu \tilde{\omega}^\mu = 0$ on-shell. The symplectic structure is defined by
\begin{equation}\label{Omega}
    \Omega(\delta_1, \delta_2; \phi, X) = \int_\Sigma d^3 y \sqrt{\gamma} {n}_\mu {\omega}^\mu(\delta_1, \delta_2; \phi, X),
\end{equation}
where $\omega^\mu = G^{-1/2} \tilde \omega ^\mu$, $\Sigma$ is a complete hypersurface with induced volume element $\sqrt{\gamma}$, and ${n}_\mu$ is the unit normal to $\Sigma$.  We emphasize that, due to the existence of a good canonical action (\ref{S scalar parametrized}) for this system, the covariant phase space defined by this symplectic structure is equivalent to the canonical phase space taken on-shell; i.e., it is equivalent to the Dirac brackets defined by the Poisson bracket and the constraints $P_\alpha + H_\alpha^\phi =0$.

The symplectic structure $\Omega$ is independent of $\Sigma$ if the symplectic flux,
\begin{equation}\label{flux}
    \Phi = \int_{\partial M^\mu} \sqrt{h} \rho_\mu {\omega}^\mu = \int_{\partial M^\mu} \bar{\rho}_\mu \tilde{\omega}^\mu
\end{equation}
\noindent vanishes. Here we have defined $\bar{\rho}_\mu dx^\mu = dr$ so that $\delta \bar{\rho}_\mu = 0$. Let us also introduce the radial lapse $N_r$ which satisfies $N_r \bar \rho_\mu = \rho_\mu$.
As noted above, the vanishing of (\ref{flux}) is guaranteed so long as we have have been careful to define $\boldsymbol \theta$ as dictated by our calculation of $\delta S$ in section \ref{PFT}.  We may check this explicitly as follows. It is straightforward to verify that
\begin{equation}\label{flux 1}
    \rho_\mu \omega^\mu = -\frac{1}{2} \delta_1 X^\alpha \delta_2 X^\mu _\alpha \rho_\mu (\nabla \phi)^2 + \rho_\mu \delta_2 \pi^\mu ( \partial_\alpha \phi \delta_1 X^\alpha - \delta_1 \phi ) - (1 \leftrightarrow 2)
\end{equation}
\noindent where we have used the boundary condition for the embedding variables $\rho_\alpha \delta X^\alpha = 0$ as well as $\rho_\alpha \theta^\alpha = 0$, which follows from the fact that all boundary terms in the variation of $S_0^P$ vanish for the stated boundary conditions. The first term in (\ref{flux 1}) vanishes under anti-symmetrization:
\begin{eqnarray*}
    \delta_1 X^\alpha \delta_2 X^\mu _\alpha \rho_\mu - (1 \leftrightarrow 2)  &=& - \rho_\beta \delta_1 X^\alpha \partial_\alpha \delta_2 X^\beta - (1 \leftrightarrow 2) = \partial_\alpha \rho_\beta \delta_1 X^\alpha \wedge \delta_2 X^\beta = \nabla_\beta \rho_\alpha \delta_1 X^\alpha \wedge \delta_2 X^\beta \\
     &=& - K_{AB} X^A_\alpha X^B_\beta \delta_1 X^\alpha \wedge \delta_2 X^\beta = 0,
\end{eqnarray*}
where we have again used $\rho_\alpha \delta X^\alpha = 0$ and the fact that the extrinsic curvature $K_{AB}$ is symmetric. The second term in (\ref{flux 1}) then vanishes for either Dirichlet boundary conditions (where $0  = \partial_\alpha \phi \delta X^\alpha$) or Neumann boundary conditions (where $\rho_\mu \delta \pi^\mu = N_r \bar \rho_\mu \delta \pi^\mu = N_r\delta (\bar  \rho^\mu \partial_\mu \phi) =0$).

Now, the canonical analysis of these boundary conditions suggested that all diffeomorphisms which preserve $\partial M$ are pure gauge, whether or not their action on $\partial M$ is trivial.  In the covariant phase space formalism, pure gauge transformations correspond to degenerate directions of the symplectic structure.  It is not hard to show that this is the case for such diffeomorphisms.  Specifically, we now verify that given a solution of the full equations of motion $(X,\phi)$, a linearized solution $(\delta_2 X, \delta_2 \phi)$ and the variation $(\delta_1 X, \delta_1 \phi)$ along a diffeomorphism as in (\ref{gauge syms}), we have
\begin{equation*}
    \Omega(\delta_1, \delta_2) = 0.
\end{equation*}

It is convenient to note that, under (\ref{gauge syms}), all the quantities transform as tensors in their internal $\mu$ indices and as scalars in the target $\alpha$ indices, e.g.
\begin{equation*}
    \delta T^\mu_\alpha = \epsilon^\nu \nabla_\nu T^\mu_\alpha - T^\nu_\alpha \nabla_\nu \epsilon^\mu .
\end{equation*}
Using this property and the fact that the background satisfies the equations of motion, it is not difficult to show that
\begin{equation}\label{omega gauge 1}
    \Omega =  \int_\Sigma d^3 x \sqrt{\gamma} n_\mu \nabla_\nu(\epsilon^\mu \theta[\delta_2]^\nu - \epsilon^\nu \theta[\delta_2]^\mu).
\end{equation}
Furthermore, given a skew-symmetric space-time tensor $A^{\mu \nu}$,
\begin{equation}\label{identity cov der}
   n_\mu \nabla_\nu A^{\mu \nu} = - D_a A^{\bot a} = D_a( n_\mu A^{\mu \nu} X^a_\nu ) ,
\end{equation}
\noindent where $D$ is the covariant derivative on the slice $\Sigma$ (i.e.,  $D_b v^a = X_\alpha^a X^\beta_b \nabla_\beta( v^c X^\alpha_c ) )$.  Moreover, for any vector field $V^a$ on $\Sigma$ Stokes' theorem allows us to write
\begin{equation*}
    \int_\Sigma \sqrt{h} D_a V^a = \int_{\partial \Sigma} \sqrt{h_\partial} \hat{\rho}_a V^a,
\end{equation*}
\noindent where $\hat \rho_a$ is the unit normal to $\partial M$ in $\Sigma$.  Since this notion of ``unit normal'' is defined with respect to the induced metric $\gamma_{ab}$, the one-form $\hat \rho_a$ differs from the pullback of $\rho_\alpha$ to $\Sigma$ by a normalization factor $\sigma$:
    $\hat{\rho}_a = \sigma X^\alpha_a \rho_\alpha$, where $\sigma = [1 + (\rho \cdot n)^2]^{-1/2}$.

Using these results, we can write (\ref{omega gauge 1}) as a boundary term on the cut $\partial \Sigma$ of $\partial M$:
\begin{equation}\label{omega gauge 2}
    \Omega = \int_{\partial \Sigma} dS \sqrt{h_\partial} \sigma \rho_a X^a_\nu n_\mu (\epsilon^\mu \theta_2^\nu - \epsilon^\nu \theta_2^\mu),
\end{equation}
Decomposing $\rho_\mu = \rho_\bot n_\mu + X^a_\mu \rho_a$, we finally obtain
\begin{equation}\label{omega gauge}
    \Omega = \int_{\partial \Sigma} dS \sqrt{h_\partial} \sigma [(\epsilon^\nu n_\nu ) (\theta_2^\mu \rho_\mu) - (\epsilon^\mu \rho_\mu ) (\theta_2^\nu n_\nu )] = 0,
\end{equation}
which vanishes since the action is differentiable $(\theta_2^\mu \rho_\mu \big |_{\partial M} = 0)$ and the diffeomorphism preserves the boundary $(\epsilon^\mu \rho_\mu \big |_{\partial M} =0)$. As a result, any generator of diffeomorphisms (including time-translations) must be a constant, and can be taken to vanish on-shell.\\

The above results hold for any of our boundary conditions on $X^\alpha$, whether minimal, velocity-fixed, or fully-fixed.  Of course, for each case we must restrict the variations considered to be compatible with the boundary conditions.  In the velocity-fixed case, one therefore allows the action of a diffeomorphism only if it preserves these boundary conditions.  But for such cases the diffeomorphism is again a degenerate direction of the symplectic structure, and this is in particular the case for time-translations.  So, again, we make take the generator to vanish.

As in the canonical analysis,
the fully-fixed boundary conditions are more subtle as they are not invariant under time-translations.  We therefore postpone their discussion to section \ref{FF}.  However, we will see there that, whenever we find a well-defined boundary Hamiltonian, it again vanishes on shell.

\subsection{Robin case}
\label{Rcov}

We now turn to the case of the Robin boundary conditions (\ref{robin bcs}) on the field $\phi$.  As usual, we begin with a discussion of minimal boundary conditions for the $X^\alpha$.

As we mentioned in section \ref{PFT}, differentiability of the action is achieved by adding the boundary term (\ref{S robin}). Following \cite{CM}, we write the associated symplectic structure as
\begin{equation}\label{omega robin}
    \Omega = \int_\Sigma \bar{n}_\mu \tilde{\omega}_{bulk}^\mu + \int_{\partial \Sigma} \bar{n}_A \tilde{\omega}^A,
\end{equation}
where $\tilde \omega_{bulk}$ is the densitized symplectic current from section \ref{DNcov},  $\tilde{\omega}^A$ is the symplectic current on the boundary defined by $S_{Robin, bndy}$ in (\ref{S robin}), considered as a densitized current with respect to the boundary metric $h_{AB}$, and $n_A$ the projection of the space-time normal along the boundary. The bars on $\bar n_\mu$, $\bar n_A$ denote the inclusion of an appropriate factor of the lapse to make the integrand a density in the relevant submanifold, i.e. $\bar{n}_\mu = N^{-1} n_\mu$ and $\bar n_A = N^{-1} n_A$.

The symplectic potential associated to the boundary term in (\ref{S robin}) is
\begin{equation}\label{theta robin}
    \tilde{\theta}^A_{bndy} = \sqrt{h} [\frac{\alpha}{2} \phi^2 + \beta \phi] X^A_\alpha \delta X^\alpha,
\end{equation}
\noindent where we have used the boundary condition (\ref{robin bcs}) to replace the derivative $\partial_\rho \phi$ with $\alpha \phi + \beta$.  Taking another variation and anti-symmetrizing, we find the boundary symplectic current
\begin{eqnarray}\label{bndy omega 1}
\nonumber
    \tilde{\omega}^A &=& \partial_B( \sqrt{h} X^B_\beta \delta_1 X^\alpha \wedge \delta_2 X^\beta X^A_\alpha) [\frac{\alpha}{2}\phi^2 + \beta \phi] \\
    &+& \sqrt{h} X^A_\alpha ( \frac{\partial_\beta \alpha}{2} \phi^2 + \beta \partial_\beta \phi  ) \delta_1 X^\alpha \wedge \delta_2 X^\beta + \sqrt{h} X^A_\alpha \partial_\rho \phi ( \delta_1 X^\alpha \delta_2 \phi - \delta_2 X^\alpha \delta_1 \phi).
\end{eqnarray}

Let us verify conservation under the new boundary conditions. This amounts to showing that
\begin{equation}\label{cons robin}
    \int_R \tilde{\omega}^\mu_{bulk} \bar{\rho}_\mu + \int^{\partial \Sigma_2}_{\partial \Sigma_1} \tilde{\omega}^A_{bndy} \bar{n}_A = 0
\end{equation}
\noindent where $R$ is a region of the time-like boundary such that $\partial R = \partial \Sigma_2 - \partial \Sigma_1 $ where $\partial \Sigma_i$ are two cuts of the boundary and $\int^{\partial \Sigma_2}_{\partial \Sigma_1} \equiv \int_{\partial \Sigma_2} - \int_{\partial \Sigma_1} $. Denoting $\Delta_i = \partial_\alpha \phi \delta_i X^\alpha - \delta_i \phi $, the first term in (\ref{cons robin}) can be written
\begin{equation}\label{flux from bulk 0}
   \int_R \tilde{\omega}^\mu_{bulk} \bar{\rho}_\mu = \int_R \partial_A \left \{ \sqrt{h} X_\alpha^A \partial_\rho \phi [ \Delta_1 \delta_2 X^\alpha - \Delta_2 \delta_1 X^\alpha ]      \right\}.
\end{equation}
Moreover, by applying Stokes' theorem\footnote{Note that the extra minus sign is correlated to the fact that the normal is time-like.}
\begin{equation*}
    \int_{R} \partial_A (\sqrt{h} V^A)  = - \int_{\partial \Sigma_2 - \partial \Sigma_1} \hat{n}_A V^A ,
\end{equation*}
\noindent we can rewrite (\ref{flux from bulk 0}) as
\begin{equation}\label{flux from bulk}
   \int_R \tilde{\omega}^\mu_{bulk} \bar{\rho}_\mu =  - \int^{\partial \Sigma_2}_{\partial \Sigma_1} \sqrt{h_\partial} \sigma n_\alpha \partial_\rho \phi [ \Delta_1 \delta_2 X^\alpha - \Delta_2 \delta_1 X^\alpha ].
\end{equation}
Let us now calculate the integrals over $\partial \Sigma_{1,2}$ in (\ref{cons robin}). Looking at (\ref{bndy omega 1}), we note that a good strategy to calculate these integrals is to integrate by parts the first term in (\ref{bndy omega 1}) and move the derivative to the factor $[\alpha \phi^2/2 +\beta \phi]$. This can easily be done noting that, for a skew-symmetric tensor $C^{AB}$,
\begin{eqnarray}\label{id ibp}
\nonumber
    \bar{n}_A \partial_B (\sqrt{h} C^{AB} ) &=& \frac{1}{N \sigma} \hat{n}_A \sqrt{h} \mathcal{D}_B C^{AB} = \sqrt{h_\partial} \hat{n}_A \mathcal{D}_B C^{AB} \\
    &=& \sqrt{h_\partial} \mathfrak{D}_I ( \hat{n}_A C^{AB} X^I_B ) = \partial_I ( \sqrt{h_\partial} \hat{n}_A C^{AB} X^I_B )
\end{eqnarray}
\noindent where used have been made of (\ref{identity cov der}) and $\sqrt{h} = N \sigma \sqrt{h_\partial}$. The index $I$ corresponds to an index in the cut $\partial \Sigma$ and the symbols $\mathcal{D}$ and $\mathfrak{D}$ denote the covariant derivatives along the boundary $\partial M$ and the cut of the boundary $\partial \Sigma$, respectively. Using the result (\ref{id ibp}), and the fact that the boundary is compact (so no extra boundary terms arise integrating by parts) we readily find that
\begin{equation}\label{flux bndy robin}
\int^{\partial \Sigma_2}_{\partial \Sigma_1} \tilde{\omega}^A_{bndy} \bar{n}_A =  \int^{\partial \Sigma_2}_{\partial \Sigma_1} \sqrt{h_\partial} \sigma n_\alpha \partial_\rho \phi [ \Delta_1 \delta_2 X^\alpha - \Delta_2 \delta_1 X^\alpha ]
\end{equation}
\noindent so that (\ref{cons robin}) holds and the symplectic structure is conserved under minimal boundary conditions for $X^\alpha$ and Robin boundary conditions for $\phi$.

Let us now verify that diffeomorphisms preserving the boundary are degenerate directions of the above symplectic structure. Taking $\delta_1$ in (\ref{flux bndy robin}) to be a diffeomorphism, the boundary contribution to $\Omega$ takes the form,
\begin{equation}\label{omega bndy diff}
    \int_{\partial \Sigma} \tilde{\omega}^A_{bndy}\bar{n}_A = \int_{\partial \Sigma} \sqrt{h_\partial} \sigma (\epsilon^\mu n_\mu) \partial_\rho \phi(\delta_2 \phi - \partial_\alpha \phi \delta_2 X^\alpha).
\end{equation}
Using the boundary condition $\rho_\alpha \delta X^\alpha = 0$, we can easily show that this exactly cancels the term $(\epsilon^\nu n_\nu ) (\theta_2^\mu \rho_\mu)$ in (\ref{omega gauge}). As a result, diffeomorphisms that satisfy $\epsilon^\mu \rho_\mu = 0$ at the boundary are again degenerate directions of the symplectic structure. \\

We now turn to velocity-fixed boundary conditions for $X^\alpha$.  Here, a priori, one might think that there is a choice of symplectic structures.  The first choice would be to use (\ref{omega robin}) with $\tilde \omega^\mu_{bulk}$ and $\tilde \omega^A$ defined precisely as above.
However, one might think that there is an alternate choice of symplectic structure defined by first using the velocity-fixed boundary conditions to remove all time-derivatives from the boundary term $S^P_{bndy}$ and only then computing the boundary symplectic structure.  Such a prescription leads to $\tilde \theta^t =0$ and thus $\tilde \omega^t =0$.  As a result, if the hypersurface $\Sigma$ is taken to be a surface of constant $t$ we have simply

\begin{equation}
\label{alt}
    \Omega_{alternate} = \int_\Sigma \bar{n}_\mu \tilde{\omega}_{bulk}^\mu
\end{equation}
with no additional boundary term.  However, the use of this symplectic structure is not in accord with the prescription of \cite{CM}.  The relevant point is that, while the action $S_{0}^P + S^P_{bndy}$ defines a good variational principle even if one first uses the boundary conditions to eliminate time-derivatives from the boundary term, the computation to show this still requires an integration by parts in time along the boundary.  Had we not first eliminated the time derivatives from $S^P_{bndy}$, this integration by parts would be implicit in (\ref{CMdefs}).  But when such time derivatives are eliminated first, it must be considered part of the prescription to define $\boldsymbol \theta_{bulk}$; i.e., the prescription of \cite{CM} would now define a new $\boldsymbol \theta_{bulk,v-fixed}$ which turns out to differ from the old $\boldsymbol \theta_{bulk}$ precisely by ${\bf d} \boldsymbol \theta_{bndy}$  as defined by (\ref{theta robin}).  Thus the full symplectic structure $\Omega$ defined by \cite{CM} from the action
$S_{0}^P + S^P_{bndy}$ is the same whether or not one first uses the boundary conditions to eliminate time-derivatives from $S^P_{bndy}$.  Indeed, one may check directly that (\ref{alt}) is not conserved, and so does not define a good covariant phase space.

As a result, we take the velocity-fixed symplectic structure to be the same as for minimal boundary conditions, adding only the restriction that variations must preserve the additional boundary conditions. It is thus clear that diffeomorphisms preserving the boundary conditions are again degenerate directions of the symplectic structure.  Time translations are pure gauge, and the Hamiltonian may be taken to vanish.  The situation for fully-fixed boundary conditions will be discussed in section \ref{FF} below.

\subsection{Fully Fixed Boundary Conditions}
\label{FF}

In the preceding sections we have seen that, for boundary conditions that preserve time-translation symmetry, the Hamiltonian in parametrized field theories acts trivially on all observables and can be taken to vanish.  As mentioned in the introduction, this statement follows from the analysis of \cite{peierls argument}, which used an approach based on the Peierls bracket.  It would clearly be desirable to extend our results to a more a general setting including boundary conditions not invariant under time-translations and which guarantee locality of the algebra of boundary observables in the sense of footnote \ref{foot}.  This is the goal of the present section.  Specifically, we consider fully fixed boundary conditions for the $X^\alpha$ in which we require the embedding fields to approach prescribed values at the boundary: $X^\alpha \big| _{\partial M} = f^\alpha$ for some functions $f^\alpha$.

Choosing boundary conditions that break time-translation symmetry means that strict time-translations will not preserve the space of solutions and cannot be generated by any function on the covariant phase space. However, this need not be the end of the story.  The situation is quite similar to the case of particle mechanics in an external time-dependent potential $V(t)$ or magnetic potential $A_i(t)$.  In that case one has a natural definition of Hamiltonian $H(t)$ despite the lack of time-translation symmetry, though $H(t)$ is not conserved.  The point is that for any particular $t_0$ the Hamiltonian $H(t_0)$ at that time is defined to act as an infinitesimal time translation {\it only} on the initial data (say, $q^i$, $p_i$) at $t_0$.  For $t \neq t_0$,  the tangent vector $\delta_{t_0} q^i(t)$ to the space of solutions describing the motion generated by $H(t_0)$ is defined from this initial data by solving the linearized equations of motion.  I.e., we generally have
\begin{equation}
\label{t0eq}
\delta_{t_0} q^i(t) = - \dot{q}^i(t) \ \ \ {\rm and} \ \ \ \delta_{t_0} p_i(t) = - \dot{p}_i(t)
\end{equation}
only for $t=t_0$.

We would like to consider a corresponding construction for our time-dependent boundary conditions. Let us first consider the case of Dirichlet boundary conditions for the scalar $\phi$.
This setting can then be formally mapped to the one above by a change of variables on field space. We need only introduce background field configurations $\bar  \phi$, $\bar X^\alpha$ that satisfy the desired boundary conditions and then rewrite the theory in terms of $\varphi : = \phi - \bar \phi$, $\Delta^\alpha : = X^\alpha - \bar X^\alpha$, which then satisfy the homogeneous time-independent boundary conditions $\varphi = 0$, $\Delta^\alpha =0$. Of course, the new action will depend on the time-dependent background fields $\bar \phi$, $\bar X^\alpha$ which play the role of the external potentials $A_i(t), V(t)$ in the above discussion.

Then one expects this procedure to yield a well-defined (time-dependent) Hamiltonian for any such background configurations $\bar \phi, \bar X^\alpha$ (without requiring them to satisfy any equations of motion).  However, since the resulting actions are not invariant under the action of diffeomorphisms on the remaining dynamical fields $\varphi, \Delta^\alpha$, these Hamiltonians will generally not reduce to boundary terms on-shell.

One may hope to overcome this last obstacle using the fact that $\bar \phi$ and $\bar X^\alpha$ are arbitrary. As such, we may introduce a family of such backgrounds $\bar \phi_s,\bar X^\alpha_s$ labeled by the parameter $s > 0$.  Away from the boundary, we can require $\bar X^\alpha_s$ to vanish and $\bar \phi_s$ to become constant in the limit $s \rightarrow 0$.  We also assume that these background are uniformly bounded, independent of $s$.  We may then hope that diffeomorphism-invariance in the bulk is restored in this limit and that the limiting Hamiltonian becomes a pure boundary term on shell.  However, this property must be checked for any particular choice of the family $\bar \phi_s,\bar X^\alpha_s$.  Below, we study those families for which $\bar \phi_s := \bar \phi(\bar X_s)$ where $\bar \phi$ is a fixed smooth function  of its argument.  This simplifies the analysis somewhat as we need only specify $\bar X_s$ in detail.  Below, we occasionally omit the subscript $s$.

We begin by choosing some vector field $\epsilon^\mu$ tangent to the boundary $\partial M^\mu$ to define our notion of time-translation.
As in the point particle discussion, the infinitesimal transformation is specified in terms of initial data at a particular time, which in the present context means on some particular hypersurface $\Sigma$. For future reference we note that, if desired, we are free to choose $\Sigma$ to depend on the solution $\phi, X^\alpha$ about which one linearized to define the infinitesimal transformation. Since we wish to specify the transformation in terms of canonical fields that satisfy time-independent boundary conditions, it is convenient to also introduce reference configurations for the conjugate momenta.  We introduce  $\bar \pi_s (\bar X_s) := \sqrt{\bar \gamma} \Pi(\bar X_s)  $, where $\Pi(\bar X_s)$ is fixed scalar functions of $\bar X_s$ and $\bar \gamma$ is the determinant of the pullback of $G_{\mu \nu}$ to $\Sigma$ evaluated on the background $\bar X_s$.  We also introduce $\bar P_{\alpha,s}(\bar X_s):= - \overline{H^\phi_\alpha}$ where the overline on the right-hand side indicates that $H^\phi_\alpha$ is to be evaluated on the background $\bar \phi(\bar X_s)$, $\bar X^\alpha_s$.   Thus our background is chosen to satisfy the constraints.    We then define the subtracted quantities $\pi_{\varphi,s} = \pi - \bar{\pi}_s$, $P^\Delta_{\alpha,s} = P_\alpha - \bar P_{\alpha,s}$ which are canonically conjugate to $\varphi$ and $\Delta$. We then take $\tilde \delta_s$ to be the transformation on the space of solutions that acts as
\begin{eqnarray}
\label{dsinit}
\tilde \delta_s \varphi |_{\Sigma} = - \epsilon^\mu \partial_\mu \varphi|_{\Sigma}, & \tilde \delta_s \pi_\varphi |_{\Sigma} = \partial_\mu( \epsilon^\mu \pi_\varphi ) |_{\Sigma}, \cr
\tilde \delta_s \Delta |_{\Sigma} =  \epsilon^\mu \partial_\mu \Delta|_{\Sigma}, & \tilde \delta_s P^\Delta_\alpha |_{\Sigma} = \partial_\mu (\epsilon^\mu P^\Delta_\alpha) |_{\Sigma}.
\end{eqnarray}
Here we have taken into account the fact that the momenta are densities on $\Sigma$. Away from the surface $\Sigma$, the variations $\tilde \delta_s$ are defined by the fact that they must solve the linearized equations of motion\footnote{This solution is not unique due to the gauge symmetry, but one may choose any family of solutions which depends smoothly on the initial data and the background fields.} with initial data (\ref{dsinit}). Although it is not reflected in our choice of notation, it is important to note that the definition of $\tilde \delta_s$ will in general depend on the choice of $\Sigma$, as (\ref{dsinit}) will generally hold only on this surface.

We are of course interested in the limiting transformation $\tilde \delta := \lim_{s \rightarrow 0} \tilde \delta_s$.  In order to handle this discontinuous transformation we define symplectic products with $\tilde \delta$ through this limit; i.e., we {\it define}
\begin{equation}
\label{limdef}
\Omega(\tilde \delta, \delta) : = \lim_{s \rightarrow 0} \Omega(\tilde \delta_s, \delta)
\end{equation}
for any tangent vector $\delta$ to the covariant phase space.  The symplectic structure (\ref{Omega}) may be written in the form
\begin{equation}\label{Omega can}
    \Omega(\delta_1, \delta_2) = \int_\Sigma -\delta_1 H^\phi_\alpha \wedge \delta_2 X^\alpha + \delta_1 \pi \wedge \delta_2 \phi = \int_\Sigma \bar{n}_\mu \delta_1 (\sqrt{G} T^\mu_\alpha) \wedge \delta_2 X^\alpha + \delta_1 \pi \wedge \delta_2 \phi
\end{equation}
\noindent where the first equality is the standard fact that the canonical symplectic structure $\Omega_{can} = \int_\Sigma \delta p \wedge \delta q$ agrees with the covariant symplectic structure when pulled back to the constraint surface (i.e., when one imposes $P_\alpha = - H^\phi_\alpha$).  The
second equality follows from $\sqrt{G} T_{\alpha \beta} \bar{n}^\beta = - H^\phi_\alpha$ and the fact that $\delta \bar{n}_\mu = 0$. In order to evaluate (\ref{Omega can}) on $\tilde \delta$, it is convenient to write (\ref{dsinit}) in terms of the original fields. Using the fact that the reference configurations are fixed functions of $\bar X^\alpha$, (\ref{dsinit}) can be written
\begin{equation}\label{split}
    \tilde{\delta}_s = \delta_\epsilon + \hat{\delta}_s
\end{equation}
\noindent where $\delta_\epsilon$ is a diffeo along $\epsilon$ and $\hat{\delta}_s$ is given on the slice $\Sigma$ by
\begin{eqnarray}
\label{dsinit hat}
\hat \delta_s \phi =  - \epsilon^\mu \partial_\mu \bar{\phi}, & \hat \delta_s \pi   =  \partial_\alpha [ \sqrt{\bar \gamma}  \bar \Pi(\bar X) \hat \delta_s X^\alpha ], \cr
\hat \delta_s X^\alpha  =  -\epsilon^\mu \partial_\mu \bar X^\alpha , & \hat \delta_s P_\alpha  =  \partial_\beta [\sqrt{\bar \gamma}\bar p_\alpha(\bar X) \hat \delta_s X^\beta].
\end{eqnarray}

Since we have already established that diffeomorphisms are null directions of $\Omega$, it remains only to compute the symplectic product $\Omega(\hat \delta_s, \delta)$ of $\hat \delta_s$ with a general on-shell variation $\delta$ using (\ref{Omega can}).  It is useful to begin by choosing a specific form of regulator.  In particular, we take $\bar X^\alpha_s = R_s(r) \bar {\bar X}^\alpha$ where
$ \bar {\bar X}^\alpha$ is a fixed, smooth, $s$-independent background that satisfies the required boundary conditions, $r$ is the coordinate introduced earlier for which the boundary of $M^\mu$ is $r=r_0$, and $R_s(r)$ is a one-parameter family of smooth uniformly bounded functions for which $R_s(r_0) =1$ and $\lim_{s \rightarrow 0} R_s (r) =0$ for $r \neq r_0$.  It is also convenient\footnote{This step merely simplifies the calculations.  The final result is independent of this choice.} to choose a time coordinate $t$ adapted to our slice $\Sigma$ such that $t$ is constant on the slice so that $\partial_r$ is tangent to $\Sigma$.

There are several types of terms in the symplectic structure.  The terms involving $\delta \pi \hat \delta_s \phi$ and $\bar{n}_\mu \delta (\sqrt{G} T^\mu_\alpha) \hat \delta_s X^\alpha$ are of the form $A_\alpha \hat \delta_s X^\alpha$ of $A \hat \delta_s \phi$ where  $A_\alpha, A$ are smooth and independent of $s$.  The integrals of such quantities vanish in the the limit $s \rightarrow 0$.  To see this, first note that for $\mu \neq r$ the derivatives $\partial_\mu \bar X^\alpha_s$, $\partial_\mu \bar \phi_s$ are of the form $R_s$ times functions uniformly bounded in $s$.  Thus the contributions of these terms vanish as $s \rightarrow 0$.  It remains to deal with the $\partial_r$ derivatives, which we will shortly integrate by parts.  But before doing so, note that we may replace $\partial_r \bar \phi$ by $\partial_r [\bar \phi - \bar \phi(0)]$, where $\bar \phi(0)$ is the constant obtained by evaluating $\bar \phi$ at the origin $\bar X^\alpha =0$.  Integrating by parts now yields bulk terms of the form $\int_\Sigma B_\alpha \bar{\bar X}^\alpha R_s$ and $\int_\Sigma B [\bar \phi - \bar \phi(0)]$ where $B_\alpha, B$ are again smooth and independent of $s$ which vanish in the limit $s \rightarrow 0$.  Finally, we address the boundary terms.  These vanish because they are proportional to $\epsilon^r$ and $\epsilon^\mu$ is tangent to the boundary.

Next consider the term in the symplectic structure involving $\hat \delta_s \pi \delta \phi$.   All terms generated by the action of $\partial_\alpha$ in  (\ref{dsinit hat}) are of the form just discussed except the term involving
\begin{equation}
\partial_\alpha \hat \delta \bar X^\alpha = - \bar X_\alpha^\mu \partial_\mu [ \epsilon^\nu \partial_\nu \bar X^\alpha_s].
\end{equation}
Unfortunately, this term is not of the form just discussed.  Indeed, we have found no way to reduce this term to a useful boundary term in the limit $s\rightarrow 0$.  Except in the case $\bar \Pi =0$ where this term vanishes, it therefore appears that the limiting Hamiltonian is not a member of the algebra of boundary observables on $\partial \Sigma$ in the sense of \cite{DM1,DM2}; i.e., that it cannot be expressed in terms of fields and their radial derivatives evaluated on $\partial \Sigma$.  One would like to find an argument to demonstrate this conclusively, but such a task is beyond the scope of this work.

Let us therefore focus the ``pure Dirichlet'' boundary condition $\phi |_{\partial M} =0$ for which the choice $\bar \Pi =0$ is clearly allowed.  This also sets $\overline H_\alpha^\phi =0$ so that the remaining $\bar{n}_\mu \hat \delta (\sqrt{G} T^\mu_\alpha) \delta X^\alpha$ term in the symplectic structure also vanishes.  Thus, in this particular case we find $\Omega (\hat \delta_s , \delta ) \rightarrow 0$ so that the limiting Hamiltonian is trivial.

At a technical level, the problems encountered above stem from the fact that the momenta are densities on the surface $\Sigma$.  An alternate strategy is thus to work only with tensors, and thus to specify the transformation $\tilde \delta $ and thus $\hat \delta$ through a choice of velocities on $\Sigma$.  Of course, such a procedure is not a priori guaranteed to yield a canonical transformation, so we will have to check explicitly whether the $s \rightarrow 0$ limit yields a Hamiltonian vector field.  We leave the associated calculations for appendix \ref{modttrans} where arguments similar to those above yield
\begin{equation}\label{fff}
    \Omega(\tilde{\delta}, \delta)
    = \int_{\partial \Sigma} \sqrt{h_\partial} \left[ (n \cdot \rho) \sigma \dot \varphi \delta \varphi  \right] .
\end{equation}
The explicit computations in appendix \ref{modttrans} are for the Robin case, though of course the pure Neumann condition is just the special case $\alpha = \beta =0$ and the general Dirichlet case can be obtained by taking $\alpha, \beta \rightarrow \infty$ with $\alpha/\beta$ fixed.

Some comments are in order. First, we note that this clearly vanishes for Dirichlet boundary conditions ($\delta \varphi|_{\partial M} =0$), and in particular for the pure Dirichlet case $\delta \phi|_{\partial M} =0$ in agreement with our discussion above. For other boundary conditions on $\phi$, (\ref{fff}) vanishes identically only if the surface $\Sigma$ is chosen to intersect the boundary orthogonally (so that $n \cdot \rho =0$).  Such a choice means that $\Sigma$ depends on the particular solution $\phi, X^\alpha$ about which we linearized to specify the infinitesimal transformation in (\ref{dsinit}), but this creates no problems.  Making such a choice, the above notion of time-translation is once again pure gauge and the Hamiltonian vanishes on shell.

On the other hand, if $n \cdot \rho \neq 0$, it turns out that the transformation $\tilde \delta$ is not a Hamiltonian vector field.  As a result, it does not define a canonical transformation and has no generator. One may see this by taking an additional variation of (\ref{fff}) and anti-symmetrizing.
For simplicity, let us consider variations $\delta_1,\delta_2$ for which $\delta_{1,2} X^\alpha =0$. Then we find
\begin{equation}\label{integrability failed}
    \delta_2 \Omega(\tilde{\delta}, \delta_1) - \delta_1 \Omega(\tilde{\delta}, \delta_2) = \int_{\partial \Sigma} \sqrt{h_\partial} \sigma (n \cdot \rho) [\delta_1 \varphi \delta_2 \dot{\varphi} - \delta_2 \varphi \delta_1 \dot{\varphi} ].
\end{equation}
Because $\varphi$ and $\dot{\varphi}$ are independent pieces of initial data,  (\ref{integrability failed}) cannot vanish in general.  Thus the equation $\Omega(\tilde \delta, \delta) = \delta H[\tilde \delta]$ that would define the Hamiltonian $H[\tilde \delta]$ is not integrable and no Hamiltonian exists.

\section{Discussion}

Despite the presence of diffeomorphism invariance, one expects that non-gravitational parametrized theories should not have Hamiltonians given by useful boundary terms.
The results of \cite{ADM2,Kuchar 1-4} clearly show that this is the case for the simplest boundary conditions.  We have verified this expectation in detail for a broader set of boundary conditions above, including Robin boundary conditions for the dynamical scalar field $\phi$ and both velocity-fixed and fully-fixed boundary conditions for the parametrization scalars $X^\alpha$. Covariant phase space techniques were used to investigate certain subtleties.  Our analysis verifies the result of \cite{peierls argument} (based on Peierls Bracket techniques) that the Hamiltonian of parametrized field theories must vanish whenever the boundary conditions are invariant under time-translations.  For all such boundary conditions studied, time translations were degenerate directions of the symplectic structure and thus pure gauge.

Even for our fully-fixed boundary conditions, for which time-translations are not symmetries, all notions of boundary Hamiltonian that we could construct vanish on-shell, so that the transformations they generate are again pure gauge. For cases that were not pure gauge either the transformation failed to be integrable, and so had no Hamiltonian generator, or for technical reasons the resulting Hamiltonian could not be shown to reduce to a boundary observable in the sense of \cite{DM1, DM2}.  It therefore appears that
the Hamiltonian of a parametrized field theory never defines a useful boundary observable. It would of course be useful to show this more definitively, and in particular to more rigorously demonstrate that Hamiltonians associated with the transformations (\ref{dsinit}) do not reduce to boundary observables when the regulator is removed.

It is interesting to ask to what extent this result is reproduced by other formalisms commonly used to compute Hamiltonians.  Consider for example the Regge-Teitelboim approach of \cite{RT}. The key ingredient of this procedure is to ask whether the vanishing Hamiltonian we obtained is differentiable under the specified boundary conditions. At least for the simple case where we have identified a good canonical Hamiltonian, we may note that the action and the (extended) Hamiltonian only differ by the terms that do not contain spatial derivatives ($\sim p \dot{q}$), and see immediately that differentiability of the action implies differentiability of the Hamiltonian. Since the former was established in section \ref{Action principle and boundary conditions}, we conclude that the corresponding Hamiltonians are differentiable and hence vanish within the Regge-Teitelboim formalism.  While
the existence of a good canonical formalism for the Robin case remains unclear, the above argument nevertheless implies that the Legendre transform of the action is differentiable.  This suggests that any Regge-Teitelboim analysis of the Robin case will again conclude that the Hamiltonian vanishes on shell.

If we try to apply the formalisms due to Abbott-Deser-Tekin \cite{abbott deser,DT} or Barnich-Brandt \cite{cov asympt charges} to our non-gravitational parametrized theories, we encounter two major difficulties. The first is related to the fact that both methods rely upon the existence of a generalized Bianchi identity.  This identity is trivial for non-gravitational parametrized theories in the following sense: for a general theory, let $\varphi$ denote the full collection of fields and write the action of diffeomorphisms as $\delta_\epsilon \varphi = R_\mu (\epsilon^\mu) $, where $\epsilon^\mu$ is the vector field that generates the diffeomorphism. The operators $R_\alpha$ are assumed to be local and linear in the gauge parameters and their derivatives. As a consequence of gauge invariance of the action, it follows that
\begin{equation}\label{d gauge S}
    0 = \delta_\epsilon S =  \int_M \frac{\delta {\cal L} }{ \delta \varphi} R_\alpha (\epsilon ^\alpha ) = \int_M  R^+_\alpha (\frac{\delta  {\cal L} }{ \delta \varphi} ) \epsilon^\alpha  ,
\end{equation}
where the Noether operator $R^+_\alpha$ is the adjoint of $R_\alpha$, defined by integrating by parts until no derivatives act on the $\epsilon^\mu$, and  $\frac{\delta  {\cal L} }{ \delta \varphi}$ is the Euler-Lagrange variation of ${\cal L}$ defined in section \ref{cov}. A direct consequence of (\ref{d gauge S}) are the generalized Bianchi identities
\begin{equation}\label{bianchi identities}
    R^+_\alpha (\frac{\delta {\cal L} }{ \delta \varphi} )    = 0  .
\end{equation}
Now, in gravitational theories or Yang-Mills theory, the gauge transformation $R_\alpha(\epsilon^\alpha)$ contains derivatives of the gauge parameters. As a consequence, the identities (\ref{bianchi identities}) correspond to linear combinations of derivatives of the equations of motion. On the other hand, in the non-gravitational parametrized field theories considered here all fields transform as scalars\footnote{At least in the covariant formalism. In the canonical formalism the Lagrange multipliers $\lambda^\alpha$ do not transform as scalars.} under diffeomorphisms ($\delta_\epsilon \phi = \epsilon^\mu \partial_\mu \phi$ and similarly for the $X$s). As a result, the objects $R^+_\alpha (\frac{\delta {\cal L} }{ \delta \phi} ) $ appearing in (\ref{bianchi identities}) are simply algebraic combinations of the equations of motion.  In the context of the Abbott-Deser-Tekin method, this lack of derivatives leads to a quantity that satisfies an algebraic relation at each point instead of a conservation law; i.e., it does not define a conserved energy for our non-gravitational theories.  In the Barnich-Brandt formalism, the fact that $R_{\alpha}$ is self-adjoint ($R_\alpha = R^+_\alpha$) for non-gravitational parametrized field theories implies that their Noether current vanishes identically and that the associated conserved quantity vanishes.

Of course, there is in addition a further difficulty in attempting to apply both the Abbott-Deser-Tekin and Barnich-Brandt formalisms.  To define an energy, both settings require the choice of some background solution with a time-translation symmetry. This is usually straightforward for general relativity or other gravitational theories, but for non-gravitational parametrized theories all configurations with Killing symmetries are singular! The point is simply that invariance of the $X^\alpha$ requires these fields to be constant along the relevant vector field.  But then the differential map $\partial_\mu X^\alpha$ is not invertible.  As a result, the inverse ``metric'' $\gamma^{ab}$ which appears in the action (\ref{param lagrangian}) is ill-defined.  It is possible that this issue could be resolved using the formalisms of \cite{AT,CDJM,pbmc} to extend the definition of non-gravitational parametrized theories to such degenerate configurations, but we will not explore this here.

Our motivation to investigate Hamiltonians for diffeomorphism-invariant theories came from studies of holography.  Because interesting forms of holography are expected to hold only for gravitational systems, systems without gravity and with local algebras of boundary observables (see footnote \ref{foot}) should generally fail to satisfy at least one of properties (i), (ii), and (iii) from the introduction.  Our detailed investigation of parametrized theories reinforces the argument from \cite{peierls argument} that Hamiltonians should vanish for diffeomorphism-invariant theories with time-translation invariant boundary specified by scalar fields.  In particular, it  suggests that this conclusion continues to hold even in the absence of time-translations invariance so long as the Hamiltonian is a well-defined boundary observable in the sense of \cite{DM1, DM2}.

Of course, this leaves open the possibility that some other diffeomorphism-invariant but non-gravitational theory might in some sense be `holographic' so long as its boundary conditions cannot be specified in terms of scalar fields.  To our knowledge, the only examples of such theories are topological field theories.
Such theories have many properties  in common with gravitational systems, including a complete lack of local observables. Furthermore, at least for 2+1 Chern-Simons theories the Wess-Zumino-Witten construction explicitly exhibits the kind of boundary unitarity discussed in \cite{DM1,DM2}. Because they have only a finite number of bulk degrees of freedom in typical settings, one would usually not use the term `holography' to describe such theories\footnote{However, there is some ambiguity in the terminology due to the fact that pure 2+1 gravity may be reformulated as a Chern-Simons theory \cite{Witten2+1}.}.  On the other hand, higher-dimensional Chern-Simons theories can have local degrees of freedom \cite{local dof CS, dyn struc CS}  and may be interesting to investigate further. See \cite{Paniak, CS hol anomaly, hol currents, Gegenberg} for work on cases where such theories admit a metric formulation.

\section*{Acknowledgements}
We thank Max Ba\~nados for useful discussions. This work was supported in part by the US National Science Foundation under grants   PHY05-55669 and PHY08-55415 and by funds from the University of California. During this work, TA was partly supported by a Fulbright-CONICYT fellowship.

\appendix

\section{Modified time translation symmetry}

\label{modttrans}

This appendix constructs a modified notion of time translation for the case of fully fixed boundary conditions for $X^\alpha$ and computes the associated symplectic products. Our transformations are similar to those of section \ref{FF}, but are specified in terms of velocities instead of momenta.  In particular,  we introduce a family of reference configurations $\bar \phi_s$, $\bar{X}_s^\alpha$ in such a way that the boundary conditions satisfied by the subtracted configurations $\varphi_s = \phi - \bar \phi_s$, $\Delta^\alpha_s = X^\alpha - \bar X^\alpha_s$ are left invariant by the transformation $\tilde \delta$ which acts as a diffeomorphism on $\varphi, \Delta^\alpha,$ and their first derivatives evaluated on a designated hypersurface $\Sigma$. Again, we take $\bar \phi_s = \bar \phi(\bar X_s)$ to be a fixed s-independent function $\bar \phi$ of $\bar X_s$ and we are interested in the limit $s \rightarrow 0$ in which $\bar X^\alpha_s$ vanishes away from the boundary (while remaining uniformly bounded).

We consider our general Robin boundary conditions for $\phi$ so that we have
\begin{equation}\label{robin ffX bcs}
 [\partial_\rho \phi - \alpha \phi - \beta] \big|_{\partial M} = 0  , \,\,\,\,\,\,\,\,\,\,\,\,\,   X^\alpha \big|_{\partial M} = f^\alpha.
\end{equation}
In particular, we impose (\ref{robin ffX bcs}) on the reference configurations $\bar{\phi}(\bar X_s)$, $\bar{X}_s^\alpha$.  The boundary conditions for the subtracted fields $\varphi_s = \phi - \bar \phi_s$, $\Delta^\alpha = X^\alpha - \bar X^\alpha_s$ then read
\begin{equation}\label{robin varphi}
  [ \partial_\rho \varphi_s - \alpha \varphi_s ]\big|_{\partial M} = 0 , \,\,\,\,\,\,\,\,\,\,\,\,\, \Delta_s^\alpha \big|_{\partial M}  = 0 .
\end{equation}
We would like (\ref{robin varphi}) to be invariant under the transformation on the space of solutions defined by
\begin{equation}\label{modif time var}
    \tilde{\delta}_s \varphi|_\Sigma = \epsilon^\mu \partial_\mu \varphi_s \,\,\,\,\,\,\,\,\,\,\,\,\, \tilde{\delta}_s \Delta^\alpha|_\Sigma =\epsilon^\mu \partial_\mu \Delta^\alpha_s ,
\end{equation}
and by the corresponding conditions on first derivatives of $\Delta^\alpha, \varphi$ for an appropriate class of backgrounds $\bar X^\alpha$.
In terms of $\phi$ and $X^\alpha$, (\ref{modif time var}) becomes
\begin{equation}\label{modif time D}
    \tilde{\delta}_s \phi = \epsilon^\mu \partial_\mu (\phi - \bar{\phi}_s)  \,\,\,\,\,\,\,\,\,\,\, \tilde{\delta}_s X^\alpha = \epsilon^\mu \partial_\mu (X^\alpha - \bar{X}_s^\alpha) .
\end{equation}
As in section \ref{FF}, we choose a coordinate $r$ such that the boundary corresponds to $r = r_0$ and introduce the regulator $R_s(r)$ through $\bar X^\alpha_s = R_s(r) \bar {\bar X}^\alpha$ where $ \bar {\bar X}^\alpha$ is a fixed, smooth, $s$-independent background that satisfies the required boundary conditions.  Appendix \ref{exists} shows that there exist such backgrounds $\bar {\bar X}^\alpha$ for which $\tilde {\delta}_s$ does indeed leave (\ref{robin ffX bcs}) invariant for any regulator $R_s$ satisfying  $R_s(r_0) =1$ and $\partial_r R_s(r_0) =0$.  Note that these conditions are satisfied by simple regulators like $R_s(r) = e^{-{(r-r_0)^2}/s}.$

We now evaluate the symplectic product $\Omega(\tilde{\delta}, \delta)$, where $\delta$ is a linearized solution that satisfies the boundary conditions. Without loss of generality we choose the $t$ coordinate to be constant on $\Sigma$ so that $\partial_r$ is tangent to $\Sigma$.  Making the usual split $\tilde \delta_s = \delta_\epsilon + \hat \delta_s$ and using the fact that diffeomorphisms ($\delta_\epsilon$) are null directions of $\Omega$ we have
\begin{equation*}
    \Omega(\tilde{\delta}_s, \delta) = \Omega(\hat{\delta}_s, \delta) = \Omega_{bulk} (\hat{\delta}_s, \delta) + \Omega_{bndy}(\hat{\delta}_s, \delta),
\end{equation*}
\noindent where the notation reflects the fact that the Robin symplectic structure contains an explicit boundary contribution given by (\ref{bndy omega 1}). As in section \ref{FF}, integrals of the form $\int_\sigma A \hat \delta_s \phi$, $\int_{\Sigma} A_\alpha \hat{\delta}_s X^\alpha$ vanish in the limit $s \rightarrow 0$ when $A, A_\alpha$ are independent of $s$.  In addition, one has
\begin{equation}\label{id2}
   \lim_{s\rightarrow 0} \int_\Sigma A_\alpha^\mu \partial_\mu \hat{\delta}_s X^\alpha = \int_{\partial \Sigma} A_\alpha^\mu \bar{\rho}_\mu \epsilon^\alpha, \ \ \
   \lim_{s\rightarrow 0} \int_\Sigma A^\mu \partial_\mu \hat{\delta}_s \phi = \int_{\partial \Sigma} A_\alpha^\mu \bar{\rho}_\mu \epsilon^\nu \partial_\nu \bar \phi.
\end{equation}
To verify the first equality, note that
\begin{equation}\label{d hat d X}
    \partial_\mu \hat{\delta}_s X^\alpha = \bar{\rho}_\mu \partial_r \hat{\delta}_s X^\alpha - \delta^A_\mu \partial_A( \epsilon^\nu \bar{\rho}_\nu \bar{\bar X}_s^\alpha) [\partial_r R_s(r)] + [\ldots]R_s(r),
\end{equation}
\noindent where terms algebraic in $R_s$ have not been written explicitly.  Since these terms are otherwise independent of $s$, their integrals vanish in the limit $s \rightarrow 0$.  Integrating the $\partial_r R_s(r)$ term by parts yields another bulk term of this form (which can therefore again be neglected) and a boundary term proportional $\partial_A(\epsilon \cdot \rho)|_{\partial M} =0$.  So the only contribution comes from the first term in (\ref{d hat d X}).  Integrating this term by parts gives yet another bulk term that vanishes as $s \rightarrow 0$ and the boundary term on the right-hand side of the first equality in (\ref{id2}). The second equality in (\ref{id2}) follows by a similar argument.  Using these identities  it is straightforward to verify that
\begin{equation}\label{Omega R bulk}
\lim_{s \rightarrow 0}     \Omega_{bulk}(\hat{\delta}_s, \delta) = \int_{\partial \Sigma} \sqrt{h_\partial} \sigma \delta \phi [ (n \cdot \epsilon) \rho^\mu \partial_\mu \phi +  (n \cdot \rho) \tilde{\delta} \phi ].
\end{equation}
Furthermore, the explicit boundary contribution is readily found to be
\begin{equation}\label{Omega R bndy}
    \lim_{s \rightarrow 0}     \Omega_{bndy}(\hat{\delta}_s, \delta) = - \int_{\partial \Sigma} \sqrt{h_\partial} \sigma \delta \phi (n \cdot \epsilon) \rho^\mu \partial_\mu \phi
\end{equation}
Thus, for Robin boundary conditions, we find
\begin{equation}\label{omega tilde d robin}
  \Omega(\tilde{\delta}, \delta) = \int_{\partial \Sigma} \sqrt{h_\partial} \sigma (n \cdot \rho) \dot \varphi \delta \varphi ,
\end{equation}
where we have used $\tilde \delta \phi = \dot{\varphi} := \epsilon^\mu \partial_\mu \varphi$ and $\delta \phi = \delta \varphi$.

\section{Existence of $\bar {\bar{X}}$}
\label{exists}

In this appendix we show that, for any Robin boundary condition, one may construct a reference configuration $\bar {\bar X}^\alpha$ so that the transformation $\tilde \delta$ of appendix \ref{modttrans} leaves this boundary condition invariant whenever $\bar X_s = R_s(r) \bar {\bar X}^\alpha$ with $R_s(r_0)=1$ and $\partial_r R_s(r_0) =0$.  To begin, we write the Robin boundary condition as $B_R = B_N - \alpha \varphi$, with
\begin{equation}\label{nbc}
    B_N = \rho^\alpha X^\mu_\alpha \partial_\mu \varphi
\end{equation}
where $\rho^\alpha \equiv \tilde \rho^\alpha/ |\tilde \rho|$ and $\tilde \rho^\alpha = g^{\alpha \beta} \partial_\beta F$. Note that  $\rho^\alpha$ describes a set of fixed functions of the $X$'s.  Since $\delta X^\alpha|_{\partial M}=0$, we also have $\delta \rho^\alpha|_{\partial M} =0$.  Below, we omit the subscript $s$ on $\bar X^\alpha$.

Now, the reference configuration $\bar {\bar X}^\alpha$ needs to be constructed such that $\tilde{\delta} B_R = 0$. In order to find the condition on $\bar {\bar X}^\alpha$ that ensures this property, we calculate $\tilde{\delta} B_R$ explicitly. Varying (\ref{nbc}) at the boundary, we obtain
\begin{eqnarray}
\label{delta nbc}
\tilde{\delta} B_N &=& \rho^\alpha (- X_\alpha^\mu \  \partial_\mu \tilde{\delta} \Delta^\beta \ X_\beta^\nu    \partial_\nu  \varphi + X_\beta^\nu  \  \partial_\nu  \tilde{\delta}\varphi )   \cr
&=& \rho^\alpha (- X_\alpha^\mu \  \partial_\mu (\epsilon^ \nu \partial_\nu [X - {\bar X}]^\beta) \ X_\beta^\lambda   \partial_\lambda  \varphi + X_\alpha^\nu  \  \partial_\nu  (\epsilon^\lambda \partial_\lambda\varphi ) ) \cr
&=& \rho^\alpha \epsilon^\nu \partial_\nu (X^\mu_\alpha \partial_\mu \varphi) + \rho_\alpha X_\alpha^\mu \partial_\mu (\epsilon^\nu \partial_\nu {\bar X}^\beta) \ X_\beta^\lambda \partial_\lambda \varphi  \cr
&=& \epsilon^\nu \partial_\nu B_N  -
(X^\mu_\alpha \partial_\mu \varphi) \epsilon^\nu \partial_\nu \rho^\alpha + \rho^\alpha X_\alpha^\mu \partial_\mu (\epsilon^\nu \partial_\nu {\bar X}^\beta) \ X_\beta^\lambda \partial_\lambda \varphi  ,
\end{eqnarray}
where in passing from the second to the third line the terms not involving ${\bar X}^\alpha$ and in which $\epsilon$ is differentiated have canceled out. Defining
$C^\beta = - \epsilon^\nu \partial_\nu \rho^\alpha + \rho^\alpha X_\alpha^\mu \partial_\mu (\epsilon^\nu \partial_\nu {\bar X}^\beta)$, we can write
\begin{eqnarray}\label{delta Rbc}
\nonumber
    \tilde{\delta} B_R &=& \tilde{\delta} B_N - \alpha \epsilon^\mu \partial_\mu \varphi \\
\nonumber
    &=& \epsilon^\mu \partial_\mu B_N - \epsilon^\mu \partial_\mu (\alpha \varphi) + (\epsilon^\mu \partial_\mu \alpha) \varphi + C^\alpha \partial_\alpha \varphi \\
    &=&  \epsilon^\mu \partial_\mu B_R + [C^\beta + \rho^\beta \epsilon^\mu \partial_\mu \ln \alpha] \partial_\beta \varphi.
\end{eqnarray}
In the last step we have used the boundary condition (\ref{robin varphi}) to write $\varphi = \alpha^{-1} \partial_\rho \varphi$. Since we take $\epsilon^\mu$ tangent to the boundary, we have $\epsilon^\nu \partial_\nu B_R \big|_{\partial M} = 0$ and we need only require
\begin{equation}
\label{Nreq}
  C^\beta + \rho^\beta \epsilon^\mu \partial_\mu \ln \alpha  =  - \epsilon^\nu \partial_\nu \rho^\beta  + \rho^\beta \epsilon^\mu \partial_\mu \ln \alpha + \rho^\alpha X_\alpha^\mu \partial_\mu (\epsilon^\nu \partial_\nu {\bar X}^\beta) =  0
\end{equation}
on the boundary $\partial M$.

We would like to use (\ref{Nreq}) to define $\bar {\bar X}^\alpha$ for a given choice of $\epsilon$.  Let us first consider the case $R_s(r) :=  1$, so that $\bar X^\alpha = \bar {\bar X}^\alpha$ everywhere.   Note that (once $\epsilon^\mu$ is given), the first two terms in (\ref{Nreq}) can be computed entirely from the boundary conditions and are otherwise independent of the background.  However, the third term contains $X_\alpha^\mu$ which depends on more than just the boundary conditions.  This means that our choice of $\bar {\bar X}^\beta$ must depend on the particular background, but this does not constitute any problem. Suppose we decide to impose (\ref{Nreq})  (with $R_s =1$) not just at the boundary, but in some finite neighborhood of the boundary (say, in some specified region ${\cal R}$ of the $X^\alpha$ coordinates).  Then, in that region, (\ref{Nreq}) is a one-dimensional linear ordinary differential equation  for $\epsilon^\nu \partial_\nu \bar{ \bar X}^\beta$.  Recalling that $\epsilon^\nu \partial_\nu \bar {\bar X}^\beta$ is given on the boundary by the boundary conditions, we see that a solution exists and is unique in the region ${\cal R}$.  Denoting the solution by $F^\beta$, we conclude that $\bar {\bar X}^\beta$ can be found by solving the linear one-dimensional ordinary differential equation:
\begin{equation}
\dot{\bar {\bar X}}^\beta := \epsilon^\nu \partial_\nu \bar {\bar X}^\beta = F^\beta.
\end{equation}
Solutions clearly exist, but are not unique.  However, we can choose one by specifying initial data for $\bar {\bar X}^\beta$ on the surface $\Sigma$ that we use to evaluate the symplectic structure. Finally, one may choose any extension of $\bar {\bar X}^\beta$ beyond ${\cal R}$ for which $\bar {\bar X}^\beta$ remains smooth and bounded and $\partial_\mu \bar {\bar X}^\beta$ remains invertible.

It remains only to verify that (\ref{Nreq}) continues to hold for this same $\bar {\bar X}^\alpha$ and general regulators $R_s$ with $R_s(r_0)=1$ and $\partial_r R_s(r_0) = 0$. To do so, we compute $\rho^\alpha X_\alpha^\mu \partial_\mu (\epsilon^\nu \partial_\nu [R_s(r) \bar {\bar X}^\beta])$ at $\partial M$.  The term involving second derivatives of $R_s$ is proportional to $\epsilon^r|_{\partial M} =0$.  The term involving first derivatives vanishes since $\partial_r R_s(r_0) =0$.  In the remaining term we use $R_s(r_0)=1$ to conclude that $\tilde \delta B_R =0$ as desired.

\section{Detailed variation of the action for the parametrized scalar}
\label{detailed variation DN}

This appendix shows explicitly that the actions $S_0^P, S_{Robin}^P$ provide good variational principles when $F(X)=0$ and $\phi$ satisfies the appropriate Dirichlet, Neumann, or Robin boundary condition on $\partial M$. Varying $S_0^P$ with respect to $X^\alpha$, it is easy to see that
\begin{equation}\label{d S DN1 X}
    \delta S_0^P = \int_M \sqrt{G} \left[ - \frac{1}{2} T_{\alpha \beta} \partial_\gamma g^{\alpha \beta} \delta X^\gamma + T^\mu_\alpha \partial_\mu \delta X^\alpha \right].
\end{equation}
Now, using that $\nabla_\alpha$ is metric compatible, we have $\partial_\gamma g^{\alpha \beta} = - (\Gamma^\alpha \hs _{\delta \gamma} g^{\delta \beta} + \Gamma^\beta \hs _{\delta \gamma} g^{\alpha \delta}) $ and integrating the second term by parts, we can write
\begin{eqnarray}\label{d S DN2 X}
\nonumber
    \delta S_0^P &=& \int_M \sqrt{G} \left[ \Gamma^{\alpha} \hs _{\beta \gamma} T^\beta_\alpha \delta X^\gamma - \nabla_\mu T^\mu _\gamma \delta X^\gamma  \right] + \int_{\partial M} \sqrt{h} \rho_\mu T^\mu _\alpha \delta X^\alpha \\
\nonumber
    &=&  \int_M \sqrt{G}\left[ \Gamma^{\alpha} \hs _{\beta \gamma} T^\beta_\alpha \delta X^\gamma - \nabla_\mu (T^\mu _\nu X^\nu_\gamma ) \delta X^\gamma  \right] + \int_{\partial M} \sqrt{h} \rho_\mu T^\mu _\alpha \delta X^\alpha \\
\nonumber
    &=&  - \int_M \sqrt{G}  \nabla_\mu T^\mu _\nu X^\nu_\alpha \delta X^\alpha + \int_{\partial M} \sqrt{h} \rho_\mu T^\mu _\alpha \delta X^\alpha .
\end{eqnarray}
In the last line, we have used
\begin{equation*}
    \Gamma^\beta \hs_{\alpha \gamma} = X^\mu_\alpha X^\beta_\nu \nabla_\mu X^\nu_\gamma .
\end{equation*}
Computing the variation with respect to $\phi$ is much simpler, the result being
\begin{equation}\label{d S DN phi}
    \delta S_0^P = \int_M \partial_\mu( \sqrt{G} G^{\mu \nu} \partial_\nu \phi ) \delta \phi - \int_{\partial M} \sqrt{h} \rho_\mu \pi^\mu \delta \phi .
\end{equation}
Hence, from (\ref{d S DN2 X}) and (\ref{d S DN phi}), we see that the equations of motion are
\begin{equation}\label{eoms}
    \Box \phi = 0 \,\,\,\,\,\,\,\,\,\,\,\,\,\,   \nabla_\mu T^{\mu \nu} = 0 .
\end{equation}
We also conclude that the on-shell variation of the action coincides with (\ref{delta S DN}), as promised. The fact that (\ref{delta S DN}) vanishes under minimal boundary conditions for the embedding variables and either Neumann or Dirichlet boundary conditions for $\phi$ was established in the paragraph after (\ref{delta S DN}) in main text.

Let us now verify that the action $S_{Robin}^P$, given by (\ref{S param robin}), is differentiable under Robin boundary conditions for $\phi$ and minimal boundary conditions for $X^\alpha$, which correspond to (\ref{robin bcs}) and $F(X)=0$, respectively. Since the bulk calculation is identical to (\ref{delta S DN}), the on-shell variation of (\ref{S param robin}) subject to $\rho_\alpha \delta X^\alpha \big|_{\partial M} = 0$, reads
\begin{eqnarray*}
	\delta S_{Robin}^P  &\approx&   \int_{\partial M} \sqrt{h} \left \{ \partial_\alpha \phi \partial_\rho \phi \delta X^\alpha - \partial_\rho \phi \delta \phi   \right\}
	+	\int_{\partial M} \left \{ \delta \sqrt{h} [ \phi \partial_\rho \phi -  \frac{\alpha}{2} \phi^2  ]       + \sqrt{h}  \delta [ \phi \partial_\rho \phi -  \frac{\alpha}{2} \phi^2  ]   \right\} \\
	&\approx&   \int_{\partial M} \sqrt{h} \left \{ \partial_\alpha \phi \partial_\rho \phi \delta X^\alpha - \partial_\alpha  [ \phi \partial_\rho \phi -  \frac{\alpha}{2} \phi^2  ] \delta X^\alpha
	-  \phi \delta \partial_\rho \phi - \frac{1}{2} \delta(\alpha \phi^2)  \right \} \\
	&\approx&   \int_{\partial M} \sqrt{h} \left \{ -\phi \partial_\alpha [ \partial_\rho \phi - \alpha \phi ] \delta X^\alpha  + \phi \delta  [ \partial_\rho \phi - \alpha \phi ]  \right \}\\
	&\approx&  \int_{\partial M} \sqrt{h} \phi \left \{ - \partial_\alpha \beta \delta X^\alpha  + \delta \beta      \right \}  = 0 ,
\end{eqnarray*}
\noindent as promised. In passing to the second line we have used the identity
\begin{equation}\label{delta sqrt h identity}
	\delta \sqrt{h} = \partial_A(\sqrt{h} \delta X^\alpha X^A_\alpha ) + K \rho_\alpha \delta X^\alpha  ,
\end{equation}
which we verify below. After performing the integration by parts over the boundary (which does not introduce extra boundary terms since all the fields are assumed to go to zero in the far past and future), we have used that $X^A_\alpha \partial_A ( \cdot ) \delta X^\alpha = \partial_\alpha ( \cdot ) \delta X^\alpha $ when $\rho_\alpha \delta X^\alpha \big|_{\partial M} = 0$ holds. Throughout the calculation we have used that both $\alpha$ and $\beta$ are fixed functions of $X^\alpha$, so
\begin{equation*}
    \delta \alpha = \partial_\lambda \alpha \delta X^\lambda \,\,\,\,\,\,\,\, , \,\,\,\,\,\,\,\,\,\,  \delta \beta = \partial_\lambda \beta \delta X^\lambda .
\end{equation*}

We now verify (\ref{delta sqrt h identity}) by calculating both sides.   Since the form  of coefficient $K$ is not relevant for our purposes, we drop all terms proportional to $\delta X^\alpha$, denoting equality up to such terms by $\sim$.  The left hand side is given by
\begin{eqnarray*}
\delta \sqrt{h}    &=& \frac{1}{2} \sqrt{h} h^{AB} \delta h_{AB} =  \frac{1}{2}h^{AB}\delta ( X^\alpha_A g_{\alpha \beta} X^\beta_B)   \\
			&=& \frac{1}{2} \sqrt{h} [  \partial_\gamma g_{\alpha \beta} \delta X^\gamma X^\alpha_A X^{\beta A} + 2 X^A_\alpha \partial_A \delta X^\alpha       ] .
\end{eqnarray*}
In order to manipulate the right hand side, it is useful to note $\partial_\alpha g_{\beta \gamma} = \rho_\alpha \partial_\rho g_{\beta \gamma} + X^A_\alpha \partial_A g_{\beta \gamma} $ and $X^A_\alpha X_{A \beta} = g_{\alpha \beta} - \rho_\alpha \rho_\beta $. An explicit calculation reveals
\begin{eqnarray*}
	\partial_A(\sqrt{h} \delta X^\alpha X^A_\alpha ) &=& \sqrt{h} \delta X^\alpha X^A_\alpha \left[  \partial_A X^\beta_B X^B_\beta +
	\frac{1}{2} X_B^\beta X^{\gamma B} \partial_A g_{\beta \gamma} \right ]  +  \sqrt{h} [ X_\alpha^A \partial_A \delta X^\alpha + \delta X^\alpha \partial_A X^A_\alpha  ] \\
	&\sim& \delta \sqrt{h} + \sqrt{h}\delta X^\alpha [    X^A_\alpha   \partial_A X^\beta_B X^B_\beta	 + \partial_A X^A_\alpha ]  =
	\delta \sqrt{h} + \sqrt{h}\delta X^\alpha [    X^A_\alpha   \partial_B X^\beta_A X^B_\beta	 + \partial_A X^A_\alpha ]  \\
	&=& \delta \sqrt{h} + \sqrt{h}\delta X^\alpha [    - X^A_\alpha  X^\beta_A \partial_B X^B_\beta	 + \partial_A X^A_\alpha ] \\
    &=& \delta \sqrt{h} + \sqrt{h}\delta X^\alpha [    - (\delta ^\beta_\alpha - \rho^\beta \rho_\alpha) \partial_B X^B_\beta	 + \partial_A X^A_\alpha ]  \\
	&\sim& \delta \sqrt{h}  ,
\end{eqnarray*}
\noindent as promised.

\end{document}